\documentclass[12pt,a4paper]{article}

\usepackage[british]{babel}

\usepackage[a4paper,top=2cm,bottom=2cm,left=2.5cm,right=2.5cm,marginparwidth=1.75cm]{geometry}

\usepackage[style=nature, backend=biber]{biblatex} 

\DeclareLanguageMapping{british}{british-apa} 
\DeclareFieldFormat[article]{volume}{\textbf{#1}} 

\usepackage{amsmath}
\usepackage{graphicx}
\usepackage[colorlinks=true, allcolors=blue]{hyperref}
\usepackage{orcidlink}
\usepackage[title]{appendix}
\usepackage{mathrsfs}
\usepackage{amsfonts}
\usepackage{booktabs} 
\usepackage{caption}  
\usepackage{threeparttable} 
\usepackage{algorithm}
\usepackage{algorithmicx}
\usepackage{algpseudocode}
\usepackage{listings}
\usepackage{enumitem}
\usepackage{chngcntr}
\usepackage{booktabs}
\usepackage{lipsum}
\usepackage{subcaption}
\usepackage{authblk}
\usepackage[T1]{fontenc}    
\usepackage{csquotes}       
\usepackage{diagbox}
\usepackage{xcolor}
\usepackage[normalem]{ulem}
\usepackage{pdfpages}

\newcommand\redsout{\bgroup\markoverwith{\textcolor{red}{\rule[0.5ex]{2pt}{0.75pt}}}\ULon}
\algdef{SE}[PARFOR]{ParFor}{EndParFor}[1]{\textbf{parfor} #1 \textbf{do}}{\textbf{end parfor}}

\usepackage{setspace}
\usepackage{titlesec}
\titleformat{\section} 
  {\normalfont\Large\bfseries}{\thesection.}{1em}{}
  
\usepackage{float}   
\usepackage{caption} 


\title{
Distributed Quantum Optimization for Large-Scale Higher-Order Problems with Dense Interactions
}

\author[1*]{Seongmin Kim}
\author[2]{Vincent R. Pascuzzi}
\author[3]{Travis S. Humble}
\author[1]{Thomas Beck}
\author[4]{Sanghyo Hwang}
\author[5*]{Tengfei Luo}
\author[4*]{Eungkyu Lee}
\author[1*]{In-Saeng Suh}

\affil[1]{\small National Center for Computational Sciences, Oak Ridge National Laboratory, Oak Ridge, TN 37830, United States}
\affil[2]{\small IBM Quantum, IBM T.J. Watson Research Center, Yorktown Heights, NY 10598, United States}
\affil[3]{\small Quantum Science Center, Oak Ridge National Laboratory, Oak Ridge, TN 37830, United States}
\affil[4]{\small Department of Electronic Engineering, Kyung Hee University, Yongin-Si, Gyeonggi-do 17104, Republic of Korea}
\affil[5]{\small Department of Aerospace and Mechanical Engineering, University of Notre Dame, Notre Dame, IN 46556, United States}

\date{}

\begin{document}
\maketitle

\begin{abstract}


Many real-world problems are naturally formulated as higher-order optimization (HUBO) tasks involving dense, multi-variable interactions, which are challenging to solve with classical methods. Quantum optimization offers a promising route, but hardware constraints and limitations to quadratic formulations have hampered their practicality. Here, we develop a distributed quantum optimization framework (DQOF) for dense, large-scale HUBO problems. DQOF assigns quantum circuits a central role in directly capturing higher-order interactions, while high-performance computing orchestrates large-scale parallelism and coordination. A clustering strategy enables wide quantum circuits without increasing depth, allowing efficient execution on near-term quantum hardware. We demonstrate high-quality solutions for HUBOs up to 500 variables within 170 seconds, significantly outperforming conventional approaches in solution quality and scalability. Applied to optical metamaterial design, DQOF efficiently discovers high-performance structures and shows that higher-order interactions are important for practical optimization problems. These results establish DQOF as a practical and scalable computational paradigm for large-scale scientific optimization.
\end{abstract}


\section{Introduction}
Quantum computing (QC), exploiting fundamental quantum principles such as superposition and entanglement, represents a transformative paradigm to address classically intractable problems \cite{daley2022practical, huang2022quantum, arute2019quantum}. Among its most promising applications are combinatorial optimization that arises in finance, network science, materials discovery, and energy technologies, where the search space grows exponentially with problem size \cite{ushijima2021multilevel, brandhofer2022benchmarking, lykov2023sampling, kim2026harnessing}. The ability of QC to explore multiple states simultaneously offers a potential pathway to achieve speedups in finding high-quality solutions for such complex optimization tasks \cite{cerezo2021variational, liu2023quantum}. To bridge the gap between current QC systems and the demanding large-scale applications, hybrid quantum–classical algorithms have gained substantial interest \cite{zhou2020quantum, fu2021quingo, du2022quantum}. Among them, quantum approximate optimization algorithm (QAOA) has emerged as a leading method for combinatorial optimization, successfully applied to problems such as Max-Cut and low autocorrelation binary sequences (LABS) \cite{farhi2014quantum, lykov2023sampling, shaydulin2024evidence}. These benchmark results provide early demonstrations of quantum utility compared to classical heuristics.

Most real-world optimization problems cannot be adequately captured by models restricted to pairwise (quadratic) interactions, as is typically assumed in quadratic unconstrained binary optimization (QUBO) formulations, which are computationally lossy transformations \cite{hwang2025higher}. In practice, higher-order unconstrained binary optimization (HUBO) problems arise naturally, where higher-order correlations (e.g., three-variable interactions) are important for the underlying physics \cite{battiston2021physics, hwang2025higher, romero2025bias}. Neglecting these higher-order terms can lead to oversimplified models, degraded predictive accuracy, and suboptimal solutions \cite{hwang2025higher}. Yet, most existing optimization methods are focused on quadratic formulations due to algorithmic and hardware architectural constraints. In addition, the dense (fully-connected) interactions characteristic of real-world optimization further limit the practicality of these methods \cite{kim2025quantum}. 

Although quantum algorithms are promising for directly handling problems involving higher-order interactions, current implementations are usually restricted to small, sparse problems \cite{bravyi2020obstacles, abbas2024challenges, boulebnane2024solving}. This limitation arises from constraints of current QC systems, including limited qubit counts, restricted circuit depths imposed by gate fidelity and decoherence, and intrinsic noise that degrades result reliability \cite{leymann2020bitter, de2021materials, khan2024quantum}. Consequently, practically applying quantum optimization to real-world optimization involving hundreds or thousands of densely interacting variables remain challenging. Distributed QAOA (DQAOA) has been proposed to mitigate these issues by leveraging high-performance computing (HPC)–QC integrated systems \cite{kim2024distributed}. However, it primarily focuses on quadratic problem structures and relies on the optimization of a single variational trajectory, making it vulnerable to complex energy landscapes with many local optima. Moreover, hardware utilization remains inefficient, with only a small fraction of available qubits actively utilized. Hence, developing quantum optimization frameworks capable of intrinsically handling dense, large-scale, higher-order optimization problems while effectively exploiting quantum hardware is a crucial step toward realizing the full potential of QC for real-world optimization tasks.

In this work, we develop a distributed quantum optimization framework (DQOF), in which multiple variational quantum circuits are executed in parallel. By leveraging HPC-QC integrated systems, DQOF extends quantum optimization to large-scale HUBO problems, where quadratic approximations are insufficient to faithfully capture complex interactions. DQOF systematically enhances both solution quality and scalability by explicitly exploiting the parallel execution of multiple variational instances. Furthermore, we introduce a clustering strategy that aggregates multiple sub-problems into a single wide quantum circuit, enabling simultaneous optimization under fixed circuit depth and thereby allowing the efficient utilization of quantum hardware resources. We demonstrate this clustering strategy on quantum hardware using the IBM Heron r2 processor, where hardware execution outperforms classical simulation for wide circuits ($\geq$ 28 qubits), validating the practical feasibility of the approach on near-term quantum devices. Through systematic benchmarking, we demonstrate that DQOF achieves significantly improved solution quality and scalability, enabling the solution of HUBOs with up to 500 decision variables within 170 seconds, substantially outperforming conventional quantum and classical approaches. We further show that DQOF is effective for real-world materials optimization, enabling efficient identification of high-performance designs. Collectively, these results demonstrate DQOF as a practical computational approach for large-scale higher-order optimization and highlight a viable pathway toward harnessing quantum computing for real-world scientific challenges.

\section{Results}
\subsection{HUBOs Representing Real-World Optimization}
Many real-world optimization problems can be formulated as higher-order combinatorial optimization tasks through machine learning (ML) surrogate modeling, which enables efficient representation of complex objective landscapes defined over discrete design spaces \cite{kitai2020designing, wilson2021machine, kim2022high}. In this work, we employ a higher-order factorization machine (FM) capable of capturing linear, quadratic, and cubic interactions among densely interacting decision variables \cite{rendle2010factorization, blondel2016higher, hwang2025higher}. When trained on representative datasets, the higher-order FM produces coefficients corresponding to self-, pairwise, and three-variable interactions, which are directly mapped to a HUBO (Figs.~\ref{fig:fig-1}a and S1). This HUBO formulation provides a compact and expressive representation of real-world optimization problems with dense, multi-variable interactions. The empirical distribution of HUBO elements is shown in Fig.~S2. In this study, we focus on three-variable interactions as representative higher-order terms. However, the framework is not limited to cubic interactions; higher-order terms (e.g., four- or five-variable interactions) can be incorporated by introducing additional multi-qubit entangling operations within the quantum circuits, enabling extension to more complex higher-order optimization problems.

When represented in the three-dimensional index space $[x, y, z]$, the HUBO matrix forms a triangular pyramid-shaped structure: elements with $x \neq y \neq z$ correspond to three-variable interaction terms, elements with two equal indices ($x = y \neq z$, $x \neq y = z$, $x = z \neq y$) encode pairwise interactions, and diagonal entries ($x = y = z$) represent self-interaction. These HUBOs are fully connected, leading to highly rugged energy landscapes due to the presence of three-variable interactions (Figs.~S2---S5). Such landscapes dramatically increase optimization difficulty, as the higher-order interactions generate numerous local optima and steep energy barriers that must be overcome to reach the global minimum.

Solving HUBO problems is nondeterministic polynomial-time (NP)-hard, with a computational cost that scales exponentially with the problem size (number of decision variables) $N$ \cite{abbas2024challenges}. Although brute-force search guarantees the ground truth solution, its time-to-solution increases exponentially with $N$ (Fig.~S6), making it infeasible for large system sizes. QAOA offers a promising alternative, showing potential to accelerate the solution of higher-order problems \cite{shaydulin2024evidence}. In principle, QAOA can map higher-order optimization problems onto quantum circuits by encoding both 2-local and 3-local terms corresponding to pairwise and three-variable interactions (Fig.~S7, Supplementary Note 1) \cite{campbell2022qaoa, kotil2025quantum}. However, the quantum circuit depth and gate count required to implement these higher-order problems grow rapidly with problem size $N$ (Fig.~S7c). While quantum hardware offers favorable scaling compared to quantum simulators (Fig.~S7d), noise and decoherence still pose significant challenges for solving moderate-sized HUBOs (e.g., $N=20$), resulting in low approximation ratios (the achieved objective value relative to the optimal value), indicating reduced solution quality (Fig.~S7e). These limitations motivate the development of distributed quantum frameworks capable of addressing dense, large-scale, higher-order optimization problems while effectively utilizing quantum hardware resources.

\subsection{DQOF for HUBOs}
DQOF enables a scalable solution of HUBO problems by executing multiple variational quantum circuits in parallel. Figure~\ref{fig:fig-1}a illustrates the overall workflow. The original HUBO problem is first decomposed into sub-HUBOs, which are simultaneously solved within an HPC/QC ecosystem. Each instance, which refers to an independent variational optimization run, solves sub-HUBOs, and the resulting sub-solutions are then aggregated to generate candidate global solutions, and the best candidate is selected as the final output (see Methods for details and the full algorithmic workflow). This procedure enables scalable exploration of complex, rugged optimization landscapes. Increasing the number of instances ($P$) improves solution quality while only modestly increasing time-to-solution, with overhead dominated by classical communication (Fig.~S8). With $P=10$, DQOF achieves an approximation ratio exceeding 0.99 for HUBOs ($N=40$), while maintaining a time-to-solution below 37 seconds (Fig.~\ref{fig:fig-1}b). 

Using one quantum device to solve only a single sub-HUBO is highly inefficient, particularly because current quantum processors often provide far more qubits than that required for an individual sub-problem \cite{ichikawa2024current}. To better utilize available quantum resources, we develop a clustering strategy that combines multiple sub-problem Hamiltonians, each corresponding to a sub-HUBO, into a single composite Hamiltonian. This combined Hamiltonian is then executed on a quantum device (e.g., IBM-Kingston or IBM-Fez), solving multiple sub-HUBOs simultaneously (Fig.~\ref{fig:fig-1}c, see Methods for more details). 

Through HPC–QC integration, this strategy maximizes quantum hardware utilization and allows DQOF to operate efficiently across many sub-problems. This approach is especially advantageous in the near-term quantum era, where circuit depth is the primary hardware limitation rather than circuit width \cite{bharti2022noisy}. Because the clustering method increases circuit width while leaving circuit depth unchanged, it aligns naturally with the architectural strengths and constraints of current quantum devices. Each sub-HUBOs contains $n$ variables, and we cluster $m$ sub-HUBOs into a single combined circuit. The corresponding circuit is stacked in parallel with a total width of $m \times n$ qubits (Fig.~S9). Importantly, increasing $m$ enlarges circuit width, while $n$ determines the depth and internal complexity of each sub-circuit. Since each sub-HUBO is solved independently within the same circuit layer, the overall circuit depth remains unchanged as $m$ increases. In contrast, directly solving a single HUBO of equivalent total width ($mn$) would require substantially deeper circuits due to its dense higher-order interactions (Fig.~\ref{fig:fig-1}d). Therefore, the clustering strategy decouples width from depth, providing a hardware-compatible pathway for scaling DQOF within the depth constraints of near-term quantum devices.

\begin{figure}[!ht]
\centering
\includegraphics[width=1.0\linewidth]{}
\caption{\label{fig:fig-1} \textbf{DQOF for dense large-scale HUBO problems.} \textbf{a}, Schematic of the DQOF workflow. The original HUBO is decomposed into sub-HUBOs, which are solved in parallel by independent optimization instances with different initializations. Sub-solutions are then aggregated to generate candidate global solutions, and the best-performing solution is selected. \textbf{b}, Approximation ratio and time-to-solution of DQOF as a function of problem size for $P=10$, demonstrating high accuracy and favorable scaling behavior. \textbf{c}, Construction of combined Hamiltonians by clustering $m$ sub-HUBOs, enabling multiple sub-problems to be solved simultaneously on a single QC system. The schematic is an example of a combined Hamiltonian formed by clustering eight sub-Hamiltonians and embedding it onto the IBM-Kingston coupling map. \textbf{d}, Comparison of circuit depth scaling: clustering increases circuit width while keeping depth constant, whereas directly encoding a full HUBO of the same circuit width requires substantially deeper circuits. 
}  
\end{figure}

\subsection{DQOF on Quantum Hardware}
We evaluate DQOF on IBM Heron r2 quantum processor by solving combined Hamiltonians generated through clustering 1 to 8 sub-HUBOs ($m=$ 1 to 8). Given that each sub-HUBO has size $n=4$, the resulting quantum circuits have widths ranging from 4 to 32 qubits. For each problem instance, we evaluate DQOF performance with $P=1$ across four execution configurations: a noiseless quantum simulator (`S'), a noisy emulator using a fake backend (`E'), a hybrid configuration using a noiseless simulator for expectation value minimization with real quantum hardware for sampling (`S+H') \cite{kotil2025quantum}, and full quantum hardware execution (`H').

To manage quantum hardware costs, we set the QAOA depth to two layers and limit classical optimization to 200 iterations with 10,000 measurement shots. Even under these constraints, the noiseless simulator achieves relatively high approximation ratios (Fig.~\ref{fig:fig-2}a). In contrast, the noisy emulator cannot handle circuit widths beyond 12 qubits because density-matrix simulation scales even more steeply (\textit{O}(4$^n$)) with qubit count \cite{patti2025augmenting}. The `S+H' configuration demonstrates that, although hardware sampling introduces measurement error, it still yields reasonable approximation ratios. Full hardware execution (`H') achieves reliable accuracy when $m \leq 4$, even under hardware noise. For larger $m$, the approximation ratio decreases due to accumulated hardware noise, shallow QAOA depth, and the limited classical optimization budget. Nevertheless, the results provide valuable insight into the feasibility of running DQOF on near-term quantum hardware, demonstrating its capability to solve multiple sub-HUBOs simultaneously.

More importantly, the time-scaling behavior differs drastically across configurations (Fig.~\ref{fig:fig-2}b). For all configurations that rely on classical computations (`S', `E', and `S+H'), the time-to-solution grows exponentially with circuit width, making wide-circuit execution impractical. This reveals an inaccessible execution regime in which circuit width, rather than depth, governs computational feasibility. This is because these simulators require state-vector or density-matrix representations, which scale exponentially with the number of qubits \cite{patti2025augmenting}. In contrast, quantum hardware exhibits execution times that remain approximately invariant with circuit width under fixed circuit depth and shot budgets. As a result, the quantum hardware demonstrates a clear speedup beyond circuit widths of 28 qubits, and it can successfully execute 32-qubit circuits that simulator-based configurations cannot handle within a practical timescale (Fig.~\ref{fig:fig-2}c). These findings highlight the emerging practical quantum utility of executing wide quantum circuits on quantum hardware, particularly for clustered higher-order optimization problems. To further assess hardware scalability, we evaluate DQOF on combined Hamiltonians by clustering 7 and 8 sub-HUBOs (Fig.~S10). Simulator-based execution (`S') should significantly reduce the number of QAOA iterations (200 to 42 for $m=7$, or 1 for $m=8$) to keep the time-to-solution comparable to hardware-based execution (`H', $\sim$3,000 s). Under these conditions, hardware achieves higher approximation ratios than the simulator, demonstrating a practical execution regime in which wide clustered circuits become accessible on quantum hardware while remaining impractical for simulators.

\subsection{Solving Large-Scale HUBOs using DQOF}
\begin{figure}[!ht]
\centering
\includegraphics[width=1.0\linewidth]{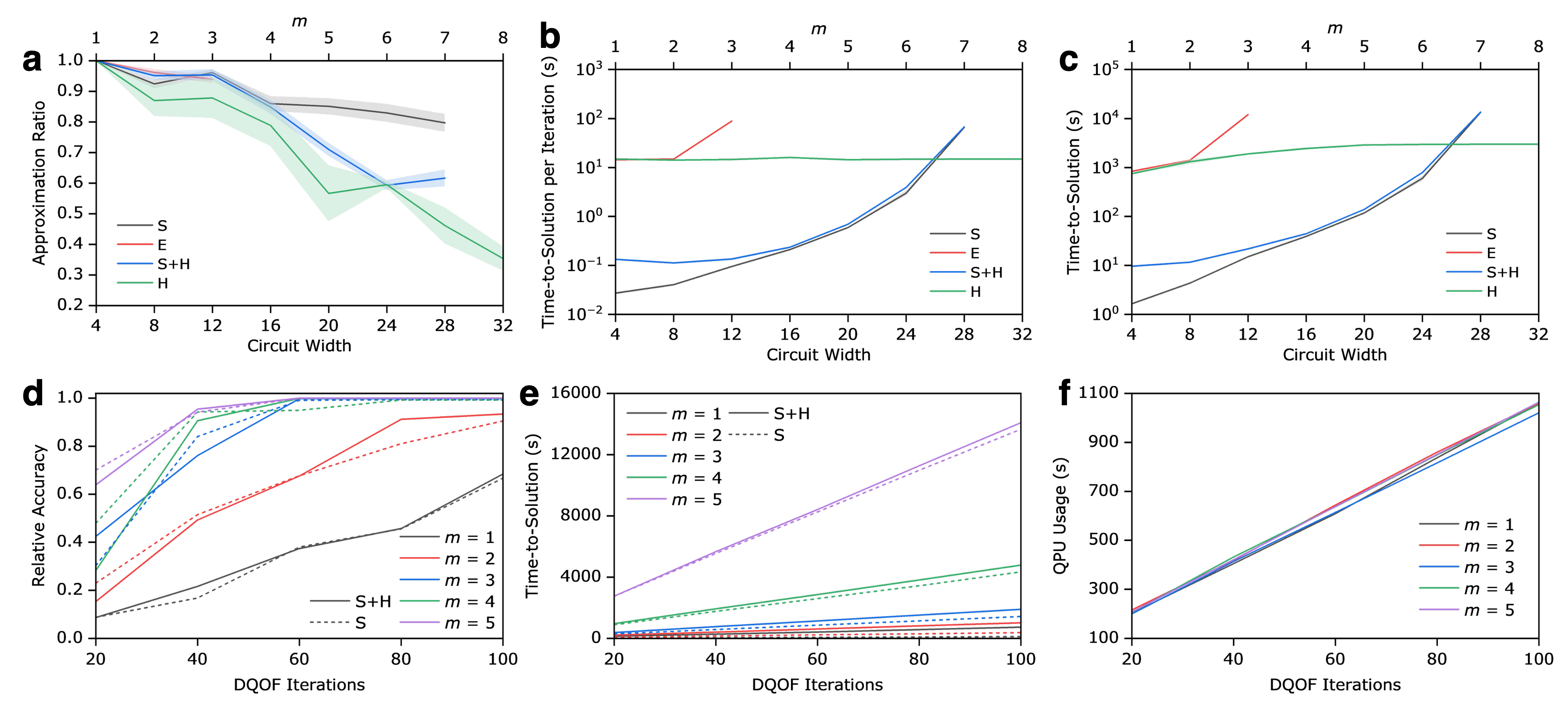}
\caption{\label{fig:fig-2} \textbf{Quantum hardware utilization and performance of DQOF.} \textbf{a}, Approximation ratios of DQOF with a single instance ($P=1$) for different execution configurations and circuit widths (i.e., different $m$). Configurations include noiseless simulator (`S'), noisy emulator (`E'), hybrid sampling mode (`S+H'), and full hardware execution (`H'). IBM quantum hardware (IBM-Kingston or IBM-Fez) and its emulator (fake backend) are used for `H' and `E' configurations. \textbf{b}, Time-to-solution per QAOA iteration and \textbf{c}, overall time-to-solution for the different configurations. Simulator-based runtimes grow exponentially with circuit width, whereas hardware execution maintains an approximately constant runtime, demonstrating hardware-enabled speedup for wide quantum circuits. \textbf{d}, Relative accuracy and \textbf{e}, time-to-solution of DQOF with clustering for a large-scale HUBO ($N=100$) as a function of DQOF iterations for different $m$. \textbf{f}, QPU usage for the `S+H' configuration with varying $m$ as a function of DQOF iterations.} 
\end{figure}

We further evaluate the performance of DQOF on a large-scale HUBO instance ($N=100$) by clustering 1 to 5 sub-HUBOs ($m=1$ to 5) and using two combined Hamiltonians ($P=2$). As expected, the relative accuracy improves with increasing DQOF iterations, and convergence becomes faster for larger $m$, because more sub-HUBOs are incorporated into each DQOF update (Fig.~\ref{fig:fig-2}d). Although the present results are obtained using a classical (`S+H') configuration, the time-to-solution scales linearly with the number of DQOF iterations (Fig.~\ref{fig:fig-2}e), while the QPU usage remains nearly constant as $m$ varies (Fig.~\ref{fig:fig-2}f). This behavior indicates that DQOF is naturally compatible with a quantum-centric supercomputing architecture, where classical HPC resources orchestrate clustering and iterative updates and quantum hardware executes the associated circuit evaluations. In such a setting, incorporating more sub-HUBOs is expected to further improve scalability and practical efficiency. These results demonstrate that DQOF, together with the clustering strategy, offers a promising scalable approach for solving dense, large-scale higher-order optimization problems and highlights the potential of quantum hardware for tackling complex HUBOs beyond the capability of existing methods.

\subsection{Performance of DQOF compared to Conventional Methods}
Notably, DQOF achieves substantially higher accuracy than conventional approaches that reduce higher-order interactions to quadratic or linear forms by introducing auxiliary variables (Supplementary Notes 2 and 3, Fig.~S11) \cite{mandal2020compressed, domino2022quadratic, bybee2023efficient, chandarana2025runtime}. Quadratization and linearization transform HUBOs into quadratic problems, enabling the use of conventional solvers \cite{mandal2020compressed, chandarana2025runtime}. However, these transformations introduce a fundamental scaling penalty. As the problem size $N$ increases, both approaches require a rapidly growing number of auxiliary variables, dramatically inflating the variable size and severely limiting scalability. Additionally, quadratization relies on finite penalty strengths to enforce auxiliary-variable constraints, and no single penalty choice is valid globally across all configurations (Fig.~S12). Thus, the original higher-order energy landscape can be distorted, leading to degraded solution quality \cite{verma2020optimal, verma2022penalty}. To compare solver performance, we evaluate relative accuracy, using the best available result as the reference baseline (Supplementary Note 4). 

\begin{figure}[!ht]
\centering
\includegraphics[width=1.0\linewidth]{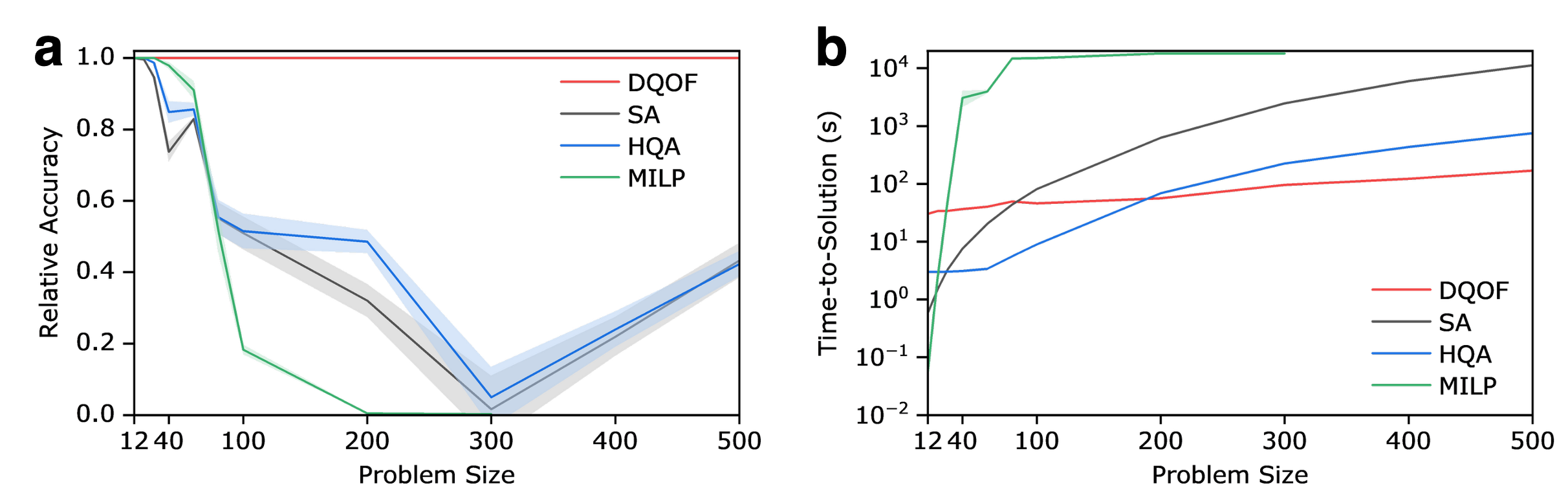}
\caption{\label{fig:fig-3} \textbf{Comparison of DQOF with classical and quantum solvers for large-scale HUBOs.} \textbf{a}, Relative accuracy and \textbf{b}, time-to-solution for different solvers. The quadratization penalty strength is fixed at 5 for SA and HQA. Conventional methods (SA and HQA based on quadratization, and MILP based on linearization) struggle to achieve the relative accuracy achieved by DQOF and incur substantially longer time-to-solution, particularly for large problems, revealing severe scalability limitations.
}
\end{figure}

Although simulated annealing (SA) and hybrid quantum annealing (HQA) can solve the quadratized QUBOs, with HQA generally exhibiting better performance \cite{kim2025quantum}, their performance is fundamentally limited by the rapid growth in variable sizes and the altered optimization landscape induced by quadratization. Consequently, solution quality degrades as $N$ increases, while the computational cost grows rapidly (Figs.~\ref{fig:fig-3}a and \ref{fig:fig-3}b). Similarly, a mixed-integer linear programming (MILP) solver (Gurobi) introduces a substantial explosion in auxiliary variables and constraints for linearizing higher-order interactions \cite{chandarana2025runtime}, causing the resulting formulations to scale poorly with $N$ and making exact solutions impractical beyond modest sizes (Figs.~\ref{fig:fig-3}a and \ref{fig:fig-3}b).

In contrast, DQOF operates directly on the native higher-order structure, avoiding auxiliary variable overhead and landscape distortion. Thus, DQOF achieves significantly higher relative accuracy with shorter time-to-solution for large-scale problems (Figs.~\ref{fig:fig-3}a and \ref{fig:fig-3}b). Its performance can be further improved by increasing the sub-HUBO size $n$ in the decomposition, thereby retaining more higher-order correlations at each iteration (Fig.~S13). Together, these results demonstrate that DQOF provides an efficient, accurate, and scalable pathway for solving dense, large-scale, and higher-order optimization problems, outperforming existing classical and quantum-based alternatives.

\subsection{Real-World Optimization using DQOF: Materials Optimization}
\begin{figure}[!ht]
\centering
\includegraphics[width=1.0\linewidth]{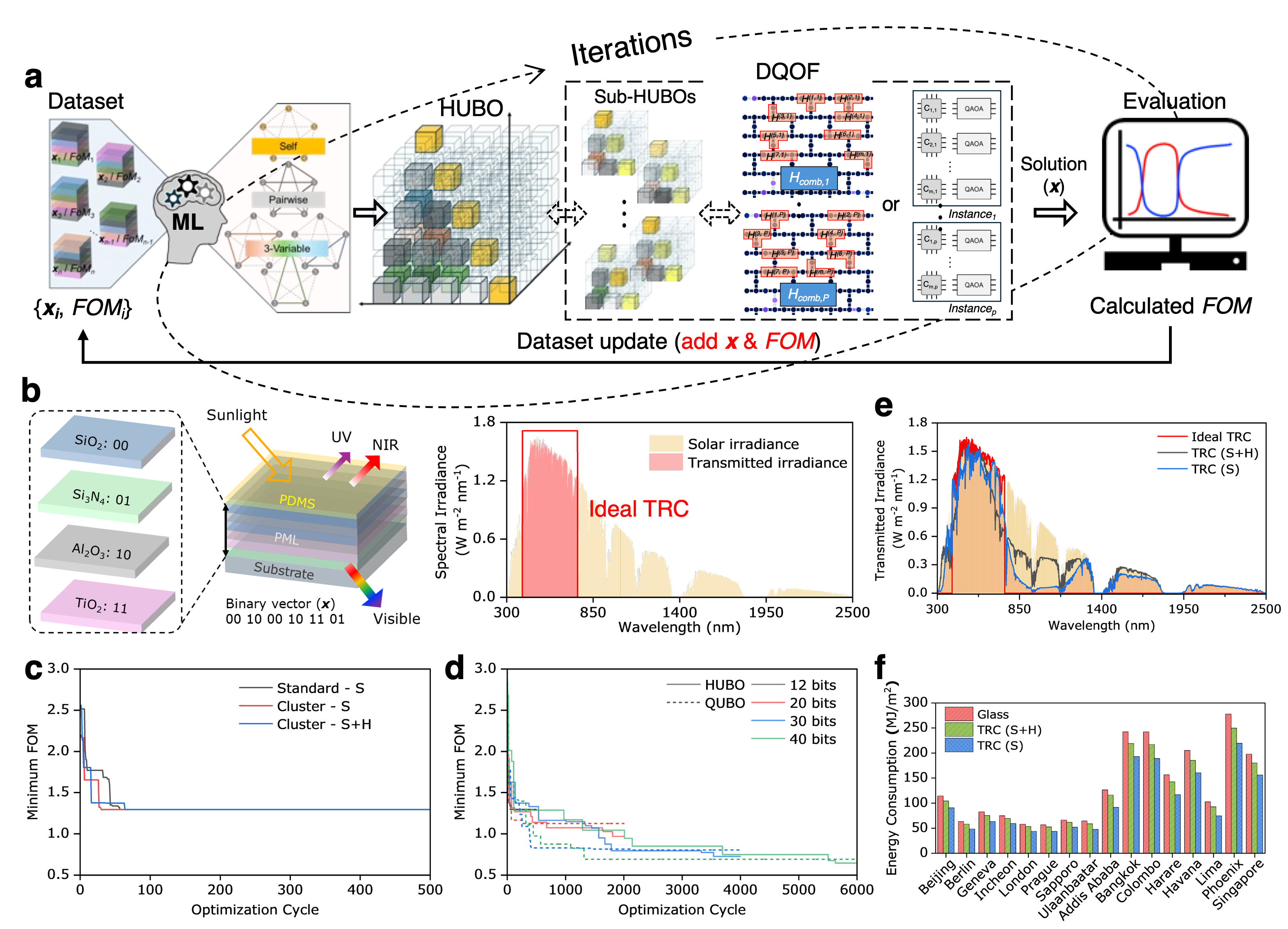}
\caption{\label{fig:fig-4} \textbf{Real-world materials optimization using AL-DQOF.} \textbf{a}, Schematic of the AL algorithm integrating higher-order FM surrogate modeling, DQOF for HUBO optimization, and physics-based simulation for data acquisition, in an iterative loop. \textbf{b}, Schematic of a TRC, consisting of a planar multilayered (PML) structure and a thermal-radiative PDMS top layer, along with the spectral irradiance of an ideal TRC within the solar spectrum. \textbf{c}, Convergence of AL-DQOF for the 12-bit system, demonstrating reliable identification of the global optimum. \textbf{d}, Convergence of AL-DQOF for the 12-, 20-, 30-, and 40-bit systems. The results highlight the importance of higher-order interactions: HUBO-based surrogates (solid line) achieve lower FOMs than QUBO-based surrogates (dotted line), showing the necessity of modeling multi-variable interactions. \textbf{e}, Transmission spectra for the ideal TRC, `S+H'-optimized TRC (12-bit system), and `S'-optimized TRC (40-bit system), showing improved spectral selectivity for larger design spaces. \textbf{f}, Cooling energy savings across international climates when replacing standard glass with the optimized TRCs, demonstrating substantial real-world energy efficiency gains enabled by AL-DQOF.}
\end{figure}

Materials optimization provides a concrete and demanding example of real-world higher-order optimization, where dense multi-variable interactions and expensive data generation pose significant challenges for conventional methods \cite{kim2022high}. Active learning (AL), which integrates ML surrogate modeling, optimization, and data generation into an iterative feedback loop, has shown strong practical utility for materials design (Fig.~\ref{fig:fig-4}a) \cite{kitai2020designing, kim2022high}. In this work, we incorporate a higher-order FM to construct surrogate models from material datasets, enabling the resulting materials optimization problems to be represented as HUBOs (Fig.~\ref{fig:fig-1}a). By using DQOF to solve these HUBOs within the AL loop, we can efficiently explore and optimize material structures to identify high-performance designs (see Methods for more details).

As a testbed to demonstrate the capabilities of AL-DQOF, we optimize layered photonic structures formed from four material choices, aiming to achieve selective transmission within the solar spectrum \cite{kim2022high, kim2024distributed}. A thin thermal radiative layer is deposited on top, resulting in a transparent radiative cooler (TRC) suitable for energy-saving window applications (Fig.~\ref{fig:fig-4}b). An ideal TRC exhibits unity transmission in the visible range and zero transmission in the ultraviolet (UV) and near-infrared (NIR) ranges. The objective of the optimization is therefore to minimize the spectral difference between the designed structure and the ideal TRC. The figure of merit (FOM) quantifies this spectral deviation, with values approaching zero indicating better-performing structures (see Methods for more details).

We first evaluate the performance of AL-DQOF with and without clustering on a small test system (12-bit case). As shown in Fig.~\ref{fig:fig-4}c, all configurations rapidly converge to the global optimal structure with a FOM of 1.295, confirming that both strategies are feasible for materials optimization (Fig.~S14). Because executing the full QAOA workflow entirely on quantum hardware within the AL loop is prohibitively costly, we use the hybrid `S+H' configuration within a quantum-centric supercomputing workflow that couples Frontier resources with two quantum devices. This configuration exhibits reliable convergence behavior and successfully discovers the global optimum, indicating that quantum hardware can be effectively incorporated into the AL-DQOF pipeline. 

The limited 12-bit design space restricts spectral selectivity, leading to relatively high FOM values for the optimized TRC. To explore higher-dimensional design spaces, we increase the system size to 40 bits and perform optimization using AL-DQOF. As expected, larger design spaces allow the discovery of higher-performing TRC, reflected in a substantially lower FOM (0.6437 for the 40-bit system, Figs.~\ref{fig:fig-4}d). Importantly, these results reveal that higher-order interactions play a critical role in materials optimization. When only linear and pairwise interactions are included, resulting in QUBO-based surrogates (dotted line), the optimized structures are consistently worse than those obtained from HUBO-based surrogates (solid line). This trend holds across all problem sizes, except for the 12-bit case, where both approaches yield identical outcomes due to the relatively small design space (Fig.~\ref{fig:fig-4}d). This strongly demonstrates that incorporating higher-order interactions is essential for achieving high-performance materials designs. 


\subsection{Performance of the Optimized Material}
Figure~\ref{fig:fig-4}e compares the transmitted irradiance of the ideal TRC, and the optimized TRCs (`S+H' configuration for the 12-bit system, and `S' configuration for the 40-bit system) across the solar spectrum. The `S+H'–optimized structure (12-bit system) exhibits the expected spectral selectivity: high transmission in the visible range and suppressed transmission in the UV and NIR ranges. The `S'-optimized TRC from the 40-bit system shows even stronger selectivity, reflected in its lower FOM, due to the larger design space that enables more effective structural optimization.

To evaluate practical benefits, we calculate the cooling energy consumption of a small office (Supplementary Note 5, Fig.~S15) \cite{li2020integration} when using standard glass, the `S+H'-optimized TRC (12-bit system), and the `S'-optimized TRC (40-bit system) as window materials. As shown in Fig.~\ref{fig:fig-4}f, the optimized TRCs substantially reduce cooling energy demand, achieving up to $\sim$28 $\%$ reduction compared to the standard glass window. These results demonstrate that AL-DQOF can be practically applied to real-world materials optimization tasks. In addition, we highlight the strong potential of integrating quantum hardware into the AL pipeline, particularly when combined with our clustering strategy, to accelerate and enhance materials design workflows.

\section{Discussion}
In this work, we develop DQOF, a distributed quantum optimization framework capable of efficiently and accurately solving dense, large-scale HUBO problems. DQOF assigns a central computational role to quantum circuits as the key for exploring and resolving higher-order optimization challenges, while classical HPC resources orchestrate large-scale parallelism, coordination, and aggregation. This architecture enables DQOF to address problem regimes that are otherwise inaccessible to either classical or quantum resources alone. 

A key feature of DQOF is its parallel sub-problem processing with quantum circuits and clustering strategy, which fundamentally alters the scaling behavior of quantum algorithms. By distributing sub-problems across HPC resources and clustering multiple sub-HUBOs into wide quantum circuits, DQOF increases effective circuit width without increasing circuit depth. This width–depth decoupling directly targets a central limitation of near-term devices, enabling hardware-compatible execution while considering higher-order structure.

Hardware experiments using IBM Heron r2 quantum processor clarify the computational role of quantum hardware in DQOF. While simulator-based executions scale exponentially with circuit width, quantum hardware maintains nearly constant runtime and successfully executes wide circuits that classical configurations cannot efficiently handle (width $\geq$ 28). This observation motivates the use of clustering to solve multiple sub-HUBOs concurrently and suggests a practical route for exploiting near-term devices by prioritizing width scaling under depth constraints. As quantum hardware improves, especially in a fault-tolerant regime, larger sub-HUBO sizes and more aggressive clustering are expected to extend the reach of this framework.

Our results show that DQOF achieves high accuracy even for HUBOs up to $N=500$ within time-to-solution under 170 seconds and substantially outperforms conventional classical and quantum approaches in which auxiliary variable overhead severely limits scalability and solution quality. By operating directly in the native higher-order optimization space, DQOF avoids these bottlenecks and maintains both accuracy and efficiency at scale. Looking forward, the performance of DQOF is expected to be further improved with GPU- and AI-enhanced strategies \cite{xu2025gpu, alexeev2025artificial}, offering an additional pathway for scaling DQOF to even larger and more complex problems. 

Beyond benchmarks, we use materials optimization as a representative scientific workflow to demonstrate how DQOF serves as a computational framework for real-world optimization tasks. By integrating higher-order FM surrogate models with DQOF, AL-DQOF efficiently navigates large, complex optimization spaces and identifies high-performance material structures. Our results further reveal that expanding the design space and explicitly incorporating higher-order interactions are essential for achieving superior material performance. Moreover, the resulting optimized TRCs achieve meaningful reductions in cooling energy consumption, illustrating the practical impact of the developed framework.

Overall, these results suggest that near-term quantum value will arise from its tight integration with large-scale classical computation. DQOF demonstrates how quantum circuits can serve as central computational primitives within a parallel and scalable optimization framework. This work highlights a path toward leveraging quantum resources as a foundational element for challenging scientific and engineering optimization tasks.

\section{Methods}
\subsection{HUBO - Representing Optimization Problems}
Higher-order terms in the HUBO representation capture cooperative interactions among multiple decision variables that cannot be faithfully represented using pairwise interactions alone. In many materials and engineering systems, the effect of one variable depends on the simultaneous configuration of several others. Therefore, the cubic terms provide an effective representation of these multi-variable correlations rather than simply enlarging the parameter space of the surrogate model. Higher-order optimization problems can be expressed in the form of HUBO, where the goal is to identify the optimal binary vector $\boldsymbol{\tilde{x}}_{\rm best}$ that minimizes the corresponding higher-order energy function $\bar{y}$ \cite{bybee2023efficient}:

\begin{equation}
\boldsymbol{\tilde{x}}_{\rm best} = \arg \min_{\boldsymbol{x}}~ \bar{y}(\boldsymbol{x}),
\label{eq:hobo_argmin}
\end{equation}

\begin{equation}
\bar{y}(\boldsymbol{x}) = \sum_{i=1}^{N} h_i x_i + \sum_{i<j} J_{ij} x_i x_j + \sum_{i<j<r} K_{ijr} x_i x_j x_r,
\label{eq:hobo_energy}
\end{equation}
where $h_i$, $J_{ij}$, and $K_{ijk}$ represent the linear, quadratic, and cubic interaction coefficients, respectively. Higher-order interactions naturally arise in many real-world applications, particularly in materials optimization, where multi-variable correlations should be accurately modeled to capture complex system behavior \cite{hwang2025higher}.

Many classical and quantum methods have been developed to solve quadratic problems. Thus, most of these approaches require a quadratization or linearization step when applied to HUBOs \cite{mandal2020compressed, domino2022quadratic, bybee2023efficient, chandarana2025runtime}. These transformations often introduce a large number of auxiliary variables, dramatically expanding the problem size and increasing landscape ruggedness, which leads to degraded solver performance. In contrast, QAOA implemented on gate-based QC can natively encode higher-order interactions into quantum circuits \cite{campbell2022qaoa}. This capability enables QAOA to address higher-order optimization tasks without relying on quadratization or linearization, offering clear advantages over conventional methods in both accuracy and scalability.

\subsection{Constructing HUBOs}
To solve general optimization problems with QC, surrogate models need to generate representations compatible with HUBOs, which capture the complex multi-variable interactions among decision variables and their associated FOMs. In this work, we employ a higher-order FM, a supervised learning model capable of learning linear, quadratic, and cubic feature interactions, to construct HUBO representations for large-scale optimization tasks (Fig.~S1) \cite{rendle2010factorization, blondel2016higher, hwang2025higher}. The higher-order FM learns the relationship between an input binary vector $\boldsymbol{x}$ and its corresponding output value $\hat{y}_{\text{3rd}}$ through a combination of multi-variable terms.

\begin{equation}
\begin{aligned}
\hat{y}_{\text{3rd}}(\boldsymbol{x})
&= b_0 + \sum_{i=1}^{N} h_i x_i
\\
&\quad
+ \frac{1}{2} \sum_{f=1}^{k}
\left[
\left( \sum_{i=1}^{N} v_{i,f} x_i \right)^2
- \sum_{i=1}^{N} v_{i,f}^2 x_i^2
\right]
\\
&\quad
+ \frac{1}{6} \sum_{f=1}^{k}
\left[
\left( \sum_{j=1}^{N} v_{j,f} x_i \right)^3
- 3 \left( \sum_{j=1}^{N} v_{j,f} x_i \right)
      \left( \sum_{j=1}^{N} v_{j,f}^2 x_i^2 \right)
+ 2 \sum_{j=1}^{N} v_{j,f}^3 x_i^3
\right],
\end{aligned}
\end{equation}
where $b_0$ denotes a global bias, $h_i$ represents linear coefficients, $v_{i,f}$ is latent factor (with rank $k$), and $x_i$ is the element of the binary input vector $\boldsymbol{x}$. Once trained on representative datasets, these coefficients are directly mapped to HUBO matrices, providing surrogates that describe how binary vector configurations influence their FOMs. The resulting HUBOs are large, dense, and contain significant higher-order interactions (Figs.~S2-S5), reflecting the complexity of real-world optimization tasks. The distribution of HUBO elements is shown in Fig.~S2, and all HUBO instances used for benchmarking are generated to follow this empirical distribution. In this study, HUBO sizes $N$, defined as the dimensionality of the binary vector encoding a material structure, range from 4 to 500 bits, enabling a comprehensive evaluation from small to practically large optimization problems.

\subsection{DQOF for Large-Scale HUBO Problems}
Consider a large-scale HUBO problem defined on $N$ binary variables $\boldsymbol{x} \in \{0,1\}^{N}$ with a problem Hamiltonian including up to 3-variable interactions:
\begin{equation}
H = \sum_{i=1}^{N} h_i x_i + \sum_{i<j} J_{ij} x_i x_j + \sum_{i<j<r} K_{ijr} x_i x_j x_r,
\end{equation}
where $h_i$, $J_{ij}$, and $K_{ijr}$ are real coefficients derived 
from the trained FM surrogate model. The following describes the workflow of DQOF.

\paragraph{1. Decomposition into sub-HUBOs.}
We define a decomposition operator $\mathcal{D}$ that partitions the global HUBO problem into $m$ sub-HUBOs. Each sub-HUBO contains $n$ variables:
\begin{equation}
\mathcal{D}(H) = \{ H^{(k)}\}_{k=1}^m, 
\end{equation}
where each sub-HUBO $H^{(k)}$ retains the 1-, 2-, and 3-variable interactions among its local variables.

\textbf{Decomposition operator via random selection.}  
Let the global variable set be 
\[
V = \{1,2,\dots,N\}.
\]
The decomposition operator $\mathcal{D}$ randomly selects $n$ variables for each sub-HUBO $k$:
\begin{equation}
V^{(k)} \subseteq V, \quad |V^{(k)}| = n, \quad V^{(k)} \sim \text{Subsets of size } n.
\end{equation}
Then the $k$-th sub-HUBO is defined as
\begin{equation}
H^{(k)} = \sum_{i \in V^{(k)}} h_i x_i
+ \sum_{\substack{i,j \in V^{(k)}\\ i<j}} J_{ij} x_i x_j
+ \sum_{\substack{i,j,r \in V^{(k)}\\ i<j<r}} K_{ijr} x_i x_j x_r,
\end{equation}
where the local variables are $x_i$ with indices $i \in V^{(k)}$. While random decomposition is used in this study, alternative strategies, such as node- or edge-based decomposition guided by interaction strength or impact factors, can also be employed to tailor sub-HUBOs to specific problem structures \cite{xu2025gpu}.

\paragraph{2. Parallel independent QAOA runs.}
Each sub-HUBO $H^{(k)}$ is independently solved using QAOA, yielding a sub-solution.
\begin{equation}
\boldsymbol{x}^{(k)} \leftarrow QAOA(H^{(k)}), 
\quad \boldsymbol{x}^{(k)} = \{x_j : j \in V^{(k)}\} \in \{0,1\}^{n}, 
\quad k=1,\dots,m.
\end{equation}

\paragraph{3. Aggregation into a global solution.}
The global solution $\boldsymbol{\tilde{x}} \in \{0,1\}^N$ is obtained through an aggregation operator $\mathcal{R}$, which determines the value of each global variable by selecting the assignment that minimizes the global HUBO energy among the potentially conflicting variables from the sub-solutions $\{\boldsymbol{x}^{(k)}\}_{k=1}^m$:
\begin{equation}
\boldsymbol{\tilde{x}} = \mathcal{R}\Big( \{ \boldsymbol{x}^{(k)} \}_{k=1}^m \Big).
\end{equation}

A simple implementation is a bitwise energy-improving update:
\begin{equation}
\tilde{x}_i =
\begin{cases}
x^{(k)}_j, & \text{if } H(\tilde{x}_1,\dots,\tilde{x}_{i-1},x^{(k)}_j,\tilde{x}_{i+1},\dots,\tilde{x}_N) < H(\boldsymbol{\tilde{x}}), \\
\tilde{x}_i, & \text{otherwise},
\end{cases}
\end{equation}
where $x^{(k)}_j$ is the variable from sub-HUBO $k$, which corresponds to the global variable $x_i$. 
When updating the global solution $\boldsymbol{\tilde{x}}$, $x^{(k)}_j$ is interpreted as a candidate assignment for $x_i$. 
Each variable from the sub-solutions is compared to the current global assignment, and the update is performed only if it decreases the global HUBO energy. In this study, the decomposition–solution–aggregation procedure is repeated for 50 iterations, with all sub-HUBOs solved in parallel at each iteration.

\paragraph{4. Parallel instance runs.}
To improve solution quality, we execute $P$ independent instances with different parameters. Each instance follows the procedure above (decomposition, parallel solution of sub-HUBOs, and aggregation), producing an independent global candidate solution
\begin{equation}
\boldsymbol{\tilde{x}}^{(p)} = \mathcal{R}\Big( \{ x^{(k,p)} \}_{k=1}^m \Big), \quad p=1,\dots,P,
\end{equation}
with corresponding global energy
\begin{equation}
E^{(p)}_{\rm global} = H(\boldsymbol{\tilde{x}}^{(p)}).
\end{equation}

\paragraph{5. Selection of the best global solution.}
Among the $P$ candidate global solutions, the final global solution $\boldsymbol{\tilde{x}}_{\rm best}$ is chosen as the one with the lowest energy:
\begin{equation}
\boldsymbol{\tilde{x}}_{\rm best} = \arg\min_{p} E^{(p)}_{\rm global}.
\end{equation}

Although sub-HUBOs are solved independently at each iteration, their coupling can be maintained through the global aggregation step, where candidate updates are evaluated against the full HUBO objective. In addition, repeated decomposition across iterations allows variables to appear in different sub-HUBOs over time, which may introduce indirect interactions between different regions of the optimization landscape.

For the complete algorithmic workflow, see Algorithm~1.
\begin{algorithm}[h!]
\caption{DQOF for HUBO Problems}\label{alg:DQOF}
\begin{algorithmic}[1]
\Require $H$ (Global HUBO), $N$ (size of the global HUBO), $m$ (Number of sub-HUBOs), $n$ (Size of sub-HUBOs), $P$ (Number of parallel runs)
\Ensure $\boldsymbol{\tilde{x}}_{\rm best}$ (Best global solution)

\ParFor{$p = 1$ to $P$} \Comment{Independent instance runs (executed in parallel)}
    \State \textbf{Initialize:} 
    \State Randomly generate a global bitstring $\boldsymbol{\tilde{x}}^{(p)} \in \{0,1\}^N$
    \\
    \State \textbf{Decomposition:}
    \For{$k = 1$ to $m$}
        \State Randomly select $n$ variables $V^{(k)} \subseteq \{1,\dots,N\}$
        \State Construct sub-HUBO $H^{(k)}$ with variables $V^{(k)}$
    \EndFor
    \\
    \State \textbf{Solution of sub-HUBOs:}
    \ParFor{$k = 1$ to $m$} \Comment{QAOA runs (executed in parallel)}
        \State $\boldsymbol{x}^{(k,p)} \gets QAOA(H^{(k)})$
    \EndParFor
    \\
    \State \textbf{Aggregation:} Update global solutions
    \For{$k = 1$ to $m$}
        \For{$j = 1$ to $n$}
            \If{$H(\tilde{x}_1^{(p)},\dots,\tilde{x}_{i-1}^{(p)}, x_j^{(k,p)}, \tilde{x}_{i+1}^{(p)},\dots,\tilde{x}_N^{(p)}) < H(\boldsymbol{\tilde{x}}^{(k,p)})$}
                \State $\tilde{x}_i^{(p)} \gets x_j^{(k,p)}$
            \EndIf
        \EndFor
    \EndFor

    \State Compute global energy $E^{(p)}_{\rm global} = H(\boldsymbol{\tilde{x}}^{(p)})$
\EndParFor
\\
\State \textbf{Selection:} Choose the best global solution
\State $\boldsymbol{\tilde{x}}_{\rm best} \gets \arg\min_{p} E^{(p)}_{\rm global}$

\State \Return $\boldsymbol{\tilde{x}}_{\rm best}$
\end{algorithmic}
\end{algorithm}

\clearpage

\subsection{Clustering Sub-HUBOs}

We consider $m$ independent sub-problems, each defined by a cost Hamiltonian $\textbf{\textit{H}}^{(k)}$ acting on $n$ qubits, for $k = 1, \dots, m$. To solve them simultaneously on a single QC system, we embed each sub-Hamiltonian into a disjoint $n$-qubit block and construct a combined cost Hamiltonian acting on $n \times m$ qubits:
\begin{equation}
\textbf{\textit{H}}_{\mathrm{comb}} = \sum_{k=1}^{m} \widetilde{\textbf{\textit{H}}}^{(k)},
\end{equation}
where the embedded Hamiltonian $\widetilde{\textbf{\textit{H}}}^{(k)}$ is defined as
\begin{equation}
\widetilde{\textbf{\textit{H}}}^{(k)} =
\left( \bigotimes_{j=1}^{k-1} I_n \right)
\otimes \textbf{\textit{H}}^{(k)}
\otimes
\left( \bigotimes_{r=k+1}^{m} I_n \right),
\label{eq:combined_hamiltonian}
\end{equation}
and $I_n$ denotes the identity operator on the $n$-qubit Hilbert space. Since each $\widetilde{\textbf{\textit{H}}}^{(k)}$ acts on disjoint qubit subsets, all terms commute, allowing the corresponding unitary evolutions to be executed in parallel. This enables simultaneous optimization of multiple sub-HUBOs within a single wide quantum circuit, improving hardware utilization and reducing overall computational latency. Figure~\ref{fig:fig-1}c illustrates the clustered Hamiltonian embedding on the IBM-Kingston device for $m=8$.

The QAOA variational state after $l$ layers is given by
\begin{equation}
|\psi(\{\boldsymbol{\beta}^{(k)}, \boldsymbol{\gamma}^{(k)}\})\rangle
=
\prod_{\ell=1}^{l}
\left[
\prod_{k=1}^{m}
e^{-i \beta^{(k)}_\ell \widetilde{\textbf{\textit{H}}}_M^{(k)}}
e^{-i \gamma^{(k)}_\ell \widetilde{\textbf{\textit{H}}}^{(k)}}
\right]
|+\rangle^{\otimes nm},
\end{equation}
where $\boldsymbol{\beta}^{(k)} = \{\beta^{(k)}_\ell\}$ and $\boldsymbol{\gamma}^{(k)} = \{\gamma^{(k)}_\ell\}$ are the variational parameters associated with the $k$-th sub-HUBO. The mixer Hamiltonian for the $k$-th block is defined as
\begin{equation}
\widetilde{\textbf{\textit{H}}}_M^{(k)} = \sum_{j \in \mathrm{block}\ k} X_j,
\end{equation}

The QAOA parameters $\{\boldsymbol{\beta}^{(k)}, \boldsymbol{\gamma}^{(k)}\}$ are optimized by minimizing the expectation value of the combined Hamiltonian:
\begin{equation}
\min_{\{\boldsymbol{\beta}^{(k)}, \boldsymbol{\gamma}^{(k)}\}}
\; \langle \psi | \textbf{\textit{H}}_{\text{comb}} | \psi \rangle.
\end{equation}

After optimization, we sample the QAOA circuit to obtain a bitstring $\boldsymbol{x}\in \{0,1\}^{nm}$. The most probable bitstring $\boldsymbol{x}^*$ is decomposed into $m$ sub-bitstrings:
\begin{equation}
\boldsymbol{x}^* = \boldsymbol{x}^{(1)} \, \| \, \boldsymbol{x}^{(2)} \, \| \, \dots \, \| \, \boldsymbol{x}^{(m)},
\end{equation}
where each $\boldsymbol{x}^{(k)} \in \{0,1\}^{n}$ corresponds to the sub-solution of the $k^{th}$ sub-problem.

Each sub-bitstring $\boldsymbol{x}^{(k)}$ is then evaluated under its corresponding cost function:
\begin{equation}
E^{(k)} = f^{(k)}(\boldsymbol{x}^{(k)}).
\end{equation}

\subsection{Computational Experiments}
All computational experiments are conducted on leadership-class HPC resources (Frontier, located at the Oak Ridge Leadership Computing Facility, unless otherwise noted. Each node consists of 64 AMD `Optimized 3rd Gen EPYC' CPU cores and 512 GB of memory). In our configuration, each core independently executes its assigned sub-HUBO while directly interfacing with a quantum backend. Up to 50 cores are employed per node. Communication across cores and nodes follows a message passing interface (MPI) workflow implemented using \texttt{mpi4py}, enabling highly concurrent execution of large numbers of QAOA at each optimization step.

For experiments, we use \texttt{qiskit} 1.4.1, \texttt{qiskit-aer} 0.15.1, \texttt{qiskit-algorithms} 0.3.1, \texttt{qiskit-optimization} 0.6.1 together with \texttt{qiskit-ibm-runtime} 0.40.1, which provides full support for interacting with IBM Quantum backends. QAOA circuits are generated using the built-in \texttt{QAOAAnsatz} and transpiled using the pass manager. Throughout all experiments, we fix the QAOA depth to two layers, number of shots to 10,000, set the transpilation optimization level to 2, initialize parameters as $\gamma=\pi/4$ and $\beta=\pi/8$, and employ COBYLA as the classical optimizer.

Quantum hardware runs are executed in a quantum-centric supercomputing setting using IBM Heron r2 processors (IBM–Kingston and IBM–Fez), selecting either both devices or the least busy device at submission time. Time-to-solution is defined as the end-to-end runtime, including decomposition, circuit execution, classical parameter updates, and aggregation, while excluding queue delays and any pre- or post-processing overhead on the quantum devices when quantum hardware is used. While simulator- and emulator-based runtimes grow exponentially with circuit width, the hardware runtime remains constant under fixed execution settings (constant circuit depth, transpiler configuration, number of shots, classical optimizer, and backend). For noise-inclusive tests, we use the emulated backend (\texttt{FakeFez}) based on the Heron r2 topology, whereas noiseless simulations rely on the \texttt{qiskit-aer} simulator.

\subsection{AL-DQOF}
Materials optimization serves as a representative and challenging testbed for demonstrating the effectiveness of DQOF on real-world higher-order optimization problems. For DQOF to reliably search the enormous materials optimization space, a high-fidelity surrogate is essential, which in turn requires sufficient training data. However, generating data in materials science is computationally expensive, and the locations of high-quality regions in the design space are unknown. AL addresses this challenge effectively by iteratively generating high-quality data points that continuously improve the surrogate’s accuracy \cite{kitai2020designing, kim2022high, kim2024distributed}. As the surrogate becomes more faithful, DQOF identifies near-optimal structures within a relatively small number of optimization cycles compared to the full design space.

The AL-DQOF consists of three key components: (1) Higher-order FM, which learns linear, quadratic, and cubic coefficients to formulate HUBO surrogates. (2) DQOF, which solves the HUBO surrogate to propose the most promising material structure. (3) Physics-based simulation, which evaluates the candidate structure’s FOM. In this study, optical properties are computed using the transfer-matrix method \cite{luce2022tmm}, a highly accurate simulation technique for multilayer optical systems.

After each iteration, the dataset is augmented with the structure proposed by DQOF and its calculated FOM. As the dataset grows, the higher-order FM generates a more accurate surrogate, enabling DQOF to identify increasingly promising candidate structures. This closed-loop workflow progressively converges toward global optimal material structures.

Because DQOF may return a structure that already exists in the dataset, we introduce a diversity-enhancing mechanism. If a repeated structure is detected, a purely random configuration (i.e., a random material design) and its corresponding FOM are added to the dataset instead. This prevents the dataset from becoming redundant and promotes broader coverage of unexplored configurations.

The AL workflow begins with an initial dataset of 25 randomly generated material structures and their FOMs. Before FM training, the dataset is split into training and test sets with a 4:1 ratio. As optimization progresses, the frequency of random insertions increases, indicating that DQOF is repeatedly returning similar structures and that convergence is being approached. Convergence is typically achieved within a few thousand optimization cycles, depending on the dimensionality of the HUBO search space. Once convergence is observed, AL-DQOF rarely discovers high-quality structures. Therefore, AL terminates when either of the following criteria is met \cite{kim2022high}:
(1) At least 90 out of the last 100 optimization cycles correspond to random insertions.
(2) The iteration count reaches predefined limits: 500, 2,000, 4,000, and 6,000 for $N=$12, 20, 30, and 40, respectively, reflecting the growing size of design spaces.

With these strategies, AL-DQOF enables efficient optimization of complex material structures. To reduce overall computational cost, the number of decomposition-solution-aggregation iterations within DQOF is set to 20.

\subsection{TRC Optimization with AL-DQOF}
TRC optimization serves as a representative material optimization for evaluating the performance of AL-DQOF. TRCs are composed of planar multilayer (PML) structures with a polydimethylsiloxane (PDMS) layer on the top for the thermal radiation through the atmospheric window \cite{lee2024integrated}. The objective is to maximize transmission in the visible regime, while suppressing transmission in the UV and NIR regimes \cite{kim2022high}. PML consists of 6 to 20 layers with a thickness of 50 nm, and each layer is chosen from four candidate materials: silicon dioxide (SiO$_2$), silicon nitride (Si$_3$N$_4$), aluminum oxide (Al$_2$O$_3$), and titanium dioxide (TiO$_2$). Each layer is encoded using a two-bit binary label (`00' for SiO$_2$, `01' for Si$_3$N$_4$, `10' for Al$_2$O$_3$, and `11' for TiO$_2$). Thus, an \textit{nl}-layer structure corresponds to a 2\textit{nl}-bit optimization problem; for instance, optimizing a 20-layer design becomes a 40-bit problem.

The performance of a designed TRC is quantified using the following FOM \cite{kim2022high}:
\begin{equation}
\text{FOM} = \frac{ 10\int_{\lambda=300}^{\lambda=2,500} [(T_{ideal}(\lambda)S(\lambda))^2 - (T_{designed}(\lambda)S(\lambda))^2]d\lambda} {\int_{\lambda=300}^{\lambda=2,500} S(\lambda)^2 d\lambda} .
\label{eqn1}
\end{equation}
where $S(\lambda)$ denotes the solar irradiance spectrum, and $T_{\mathrm{ideal}}(\lambda)$ and $T_{\mathrm{designed}}(\lambda)$ represent the transmission spectra of the ideal and designed TRC structures, respectively. $T(\lambda)S(\lambda)$ represents transmitted irradiance. Since the FOM decreases as the designed TRC more closely matches the ideal one, the optimization problem is formulated as a minimization task.

\section{Acknowledgments}
This research used resources of the Oak Ridge Leadership Computing Facility at the Oak Ridge National Laboratory, which is supported by the Office of Science of the U.S. Department of Energy under Contract No. DE-AC05-00OR22725. This material is based upon work supported by the U.S. Department of Energy, Office of Science, National Quantum Information Science Research Centers, Quantum Science Center.
{\it Notice}: This manuscript has in part been authored by UT-Battelle, LLC under Contract No. DE-AC05-00OR22725 with the U.S. Department of Energy. The United States Government retains and the publisher, by accepting the article for publication, acknowledges that the U.S. Government retains a non-exclusive, paid up, irrevocable, world-wide license to publish or reproduce the published form of the manuscript, or allow others to do so, for U.S. Government purposes. The Department of Energy will provide public access to these results of federally sponsored research in accordance with the DOE Public Access Plan (http://energy.gov/downloads/doe-publicaccess-plan).

\section*{Data availability}
The data used for this work can be found at https://github.com/Seongmin-Kim-ornl/DQOF. 

\section*{Code availability}
The codes developed for this work are available at https://github.com/Seongmin-Kim-ornl/DQOF.

\section*{Author information}
Authors and Affiliations:

\noindent
{\bf National Center for Computational Sciences, Oak Ridge National Laboratory, Oak Ridge, Tennessee 37830, United States.}\\
\noindent
Seongmin Kim, Thomas Beck \& In-Saeng Suh

\noindent
{\bf IBM Quantum, IBM T.J. Watson Research Center, Yorktown Heights, New York 10598, United States.}\\
\noindent
Vincent R. Pascuzzi

\noindent
{\bf Quantum Science Center, Oak Ridge National Laboratory, Oak Ridge, TN 37830, United States.}\\
\noindent
Travis S. Humble

\noindent
{\bf Department of Electronic Engineering, Kyung Hee University, Yongin-Si, Gyeonggi-do 17104, Republic of Korea}\\
\noindent
Sanghyo Hwang, \& Eungkyu Lee

\noindent
{\bf  Department of Aerospace and Mechanical Engineering, University of Notre Dame, Notre Dame, Indiana 46556, United States}\\
\noindent
Tengfei Luo

\section*{Contributions}
S.K. conceived the idea in discussion with I.S., E.L. and T.L. S.K. performed all experiments, developed the codes, derived the experimental results, generated the figures, conducted data analyses, and coordinated the project. All authors discussed the results, and contributed to the writing of the manuscript. 

\section*{Corresponding authors}
Correspondence to Seongmin Kim, In-Saeng Suh, Eungkyu Lee, and Tengfei Luo.

\noindent 
Email address: Seongmin Kim (kims@ornl.gov), In-Saeng Suh (suhi@ornl.gov), Eungkyu Lee (eleest@khu.ac.kr), and Tengfei Luo (tluo@nd.edu).

\section*{Ethics declarations}
\textbf{Competing Interests}\\
The authors declare no competing interests.

\printbibliography



\section*{Supplementary material}

# Supplementary Information

**Supplementary Note 1. Quantum Approximate Optimization Algorithm**

In many real-world optimization tasks, linear, quadratic, and cubic interaction terms naturally arise. We employ the quantum approximate optimization algorithm (QAOA) to solve such higher-order unconstrained binary optimization (HUBO) problems, where the cost function includes interactions beyond quadratic order. These higher-order interactions are encoded directly in the cost Hamiltonian prior to applying QAOA. A cost Hamiltonian for a HUBO can be expressed as [1]:

$$\boldsymbol{H}_C = \sum_i h_i Z_i + \sum_{i<j} J_{ij} Z_i Z_j + \sum_{i<j<k} K_{ijk} Z_i Z_j Z_k, \qquad (1)$$

QAOA is a hybrid quantum–classical variational algorithm that alternates between unitaries generated by the cost Hamiltonian and a mixer Hamiltonian to approximate the ground state of $\boldsymbol{H}_C$ [2, 3]. Classical binary variables are encoded using eigenvalues of Pauli-$Z$ operators, which enables mapping of the objective into quantum operators. This allows higher-order binary optimization problems to be expressed natively within the QAOA framework.

The algorithm begins with the initial state

$$|\psi_0\rangle = |+\rangle^{\otimes N}, \qquad (2)$$

a uniform superposition prepared by Hadamard gates. QAOA uses an alternating sequence of unitary evolutions generated by the cost Hamiltonian $\boldsymbol{H}_C$ and the mixer Hamiltonian

$$\boldsymbol{H}_M = \sum_i X_i. \qquad (3)$$

The resulting parameterized quantum state is

$$|\psi(\boldsymbol{\beta},\boldsymbol{\gamma})\rangle = e^{-i\beta_l \boldsymbol{H}_M} e^{-i\gamma_l \boldsymbol{H}_C} \cdots e^{-i\beta_1 \boldsymbol{H}_M} e^{-i\gamma_1 \boldsymbol{H}_C} |\psi_0\rangle, \qquad (4)$$

where $\boldsymbol{\beta} = (\beta_1, \ldots, \beta_l)$ and $\boldsymbol{\gamma} = (\gamma_1, \ldots, \gamma_l)$ are the variational parameters, and $l$ is the number QAOA layers.

The optimal parameters are obtained by minimizing the expectation value of the cost Hamiltonian:

$$f(\boldsymbol{\beta},\boldsymbol{\gamma}) = \langle \psi(\boldsymbol{\beta},\boldsymbol{\gamma}) | \boldsymbol{H}_C | \psi(\boldsymbol{\beta},\boldsymbol{\gamma}) \rangle. \qquad (5)$$

In principle, as $l \to \infty$, QAOA converges to the exact ground state [4]. However, increasing depth also increases classical optimization difficulty and hardware noise sensitivity. Therefore, practical QAOA implementations typically restrict $l$ to remain shallow [5, 6].

In our study, two-layer QAOA ($l = 2$) is applied to HUBO Hamiltonians. While QAOA can reliably approximate ground states for small systems (Fig. S7e), its performance

deteriorates sharply with increasing problem size.

To evaluate QAOA on HUBO instances, we use a quantum computing software development kit, Qiskit (`qiskit` v1.4.1) [7]. Quantum circuits are executed using the `qiskit-aer` simulator (v0.15.1) in *automatic* method.

**Supplementary Note 2. Quadratization Method**

A quadratic unconstrained binary optimization (QUBO) problem evaluates a binary vector $\boldsymbol{x} \in \{0,1\}^N$ as

$$f(\mathbf{x}) = \boldsymbol{x}^\mathsf{T} Q^{\text{2nd}} \boldsymbol{x} = \sum_{i=1}^{N} \sum_{j=1}^{N} Q_{ij}^{\text{2nd}} x_i x_j, \tag{6}$$

where $N$ is the problem size and $Q^{\text{2nd}}$ is a QUBO matrix encoding linear and pairwise interaction coefficients. Because the QUBO formulation is limited to quadratic terms, higher-order (cubic) interactions cannot be represented directly and should be reduced using auxiliary variables [8–10].

Consider a cubic interaction term $v_{1,2,3} x_1 x_2 x_3$. Introducing an auxiliary binary variable $x_{n+1} = x_1 x_2$ allows the cubic term to be rewritten in quadratic form as

$$v_{1,2,3} x_1 x_2 x_3 \;\to\; v_{1,2,3} x_{N+1} x_3. \tag{7}$$

This transformation increases the number of variables and expands the QUBO matrix, requiring additional qubits compared to the original higher-order unconstrained binary optimization (HUBO) formulation.

The relation $x_{N+1} = x_1 x_2$ is enforced by augmenting the objective with a penalty term weighted by a positive strength parameter $M$:

$$\min_{x_{N+1}} \left( v_{1,2,3} x_{N+1} x_3 + M\, P(x_1, x_2, x_{N+1}) \right), \tag{8}$$

where the penalty function is defined as

$$P(x_1, x_2, x_{N+1}) = 3 x_{N+1} + x_1 x_2 - 2 x_1 x_{N+1} - 2 x_2 x_{N+1}. \tag{9}$$

For binary variables, $P(x_1, x_2, x_{N+1}) = 0$ if $x_{N+1} = x_1 x_2$, and is strictly positive otherwise, ensuring that the auxiliary variable correctly represents the product at the optimum.

Applying this penalty-based quadratization procedure to all cubic interaction terms converts the original HUBO formulation into an extended QUBO representation. For a fully connected 3rd-order HUBO with problem size $N$, the number of cubic terms scales as $\binom{N}{3} = O(N^3)$. Quadratization introduces auxiliary variables to represent products of binary variables, and the number of added variables scales as $O(N^2)$, bounded by $\binom{N}{2}$. As a result, the total variable count of the quadratized problem scales as $O(N^2)$ rather than $O(N)$ for the original HUBO.

This increase in variable count has an important consequence for quantum implementation. A native 3rd-order HUBO formulation requires $O(N)$ qubits. After quadratization, however, the qubit count inflates to $O(N^2)$ because of the auxiliary variables (Fig. S12), due to the replacement of each cubic term by quadratic couplings and penalty terms. In addition, this approach critically depends on the choice of penalty strength. No single penalty value is universally valid across different HUBO instances, as the appropriate magnitude depends on the relative scale and structure of the original interaction coefficients. If the penalty is too small, constraint violations may persist, yielding incorrect solutions; if it is too large, the energy landscape becomes distorted, increasing ruggedness and impairing optimization [11, 12]. Consequently, identifying a suitable penalty strength is nontrivial and often requires problem-specific tuning, which significantly limits the robustness and scalability of quadratization-based approaches.

**Supplementary Note 3. Classical and Quantum Solvers for HUBO Problems**
To benchmark our quantum optimization framework (DQOF) against conventional approaches, we evaluate representative classical and quantum solvers for HUBO problems. Hybrid quantum annealing (HQA) and simulated annealing (SA) with quadratization are implemented using the `dimod` module from the `dwave-system` (version 1.30.0) library. All experiments use default solver settings. Classical preprocessing steps, including quadratization, are performed on the Frontier supercomputer (Oak Ridge Leadership Computing Facility), where each compute node is equipped with 64 AMD Optimized 3rd Gen EPYC CPU cores and 512 GB of memory. To examine the sensitivity of solver performance to constraint enforcement, quadratization penalty strengths are varied across values of 1, 2, 5, 10, and 30. The results show that no single penalty value consistently yields optimal performance across different problem sizes (Fig. S12).

In addition, HUBO instances are solved using the Gurobi optimizer (version 12.0.3) by reformulating the higher-order binary objective into an equivalent mixed-integer linear program (MILP) via exact linearization. Higher-order interaction terms are replaced by auxiliary binary variables, with their logical relationships enforced through linear constraints, resulting in an exact, penalty-free reformulation. The resulting MILP is solved using Gurobi to search for high-quality solutions. A time limit of five hours is imposed to bound computational cost, and the best incumbent solution is recorded via solver callbacks to accurately measure time-to-solution. All solver settings are kept at default values. MILP experiments are conducted on a workstation equipped with an AMD Ryzen Threadripper PRO 3975WX processor (32 cores at 3.5 GHz) and 32 GB of memory. While this MILP can solve small HUBO instances exactly within the imposed time limit, its performance deteriorates rapidly as the problem size increases. The time-to-solution grows exponentially with problem size, and the optimality gap exceeds 20,186,103 % for $N = 300$ even after the imposed five-hour time limit. For even larger instances (size $\geq 400$), the solver fails to complete due to memory limitations. These results highlight the severe

scalability limitations of exact linearization–based classical methods for dense, large-scale higher-order problems.

**Supplementary Note 4. Performance Analysis**

We evaluate the performance of the solvers using approximation ratios, relative accuracy, and time-to-solution across HUBO instances. Ground truth solutions are obtained using brute-force search. Because brute-force search scales exponentially with problem size, middle-level classical resources cannot feasibly compute exact solutions. Therefore, we use the Frontier supercomputer with up to $2^{14}$ CPU cores across 328 compute nodes.

The approximation ratio compares the solver's energy to the exact (brute-force) solution:

$$\text{approximation ratio} = \frac{E_\text{solver}}{E_\text{exact}}. \tag{10}$$

Even with leadership-class high-performance computing resources, brute-force becomes impractical for problems larger than 40 qubits due to the exponential growth of the search space. For larger HUBO instances, we therefore adopt a relative accuracy metric, comparing a solver's energy to the best energy achieved by any solver in this study [13]:

$$\text{relative accuracy} = \frac{E_\text{solver}}{E_\text{best}}. \tag{11}$$

We also record the time-to-solution, defined as the total wall-clock time from the start of algorithm execution to obtaining the solver's solution. These metrics characterize the scalability and practical effectiveness of DQOF for large-scale HUBO problems.

**Supplementary Note 5. Energy Savings Estimation**

To quantify the practical energy-saving benefits enabled by the optimized transparent radiative cooler (TRC) structures, we perform building cooling simulations using EnergyPlus (v9.4). The simulated domain corresponds to a standard office room with dimensions of 6 m (width) × 8 m (length) × 2.7 m (height), and two 3 m × 2 m windows. Simulations compare conventional windows to windows equipped with the optimized TRC structures (Fig. S14). The cooling setpoint is maintained at 24°C, and default EnergyPlus parameters are used except for optical inputs derived from each designed TRC.

Optical properties, including solar transmittance, solar reflectance, visible transmittance, visible reflectance, infrared transmittance, and hemispherical emittance, are obtained via transfer-matrix-method (TMM) calculations [14].

Cooling energy consumption is evaluated for sixteen international cities representing temperate, arid, and tropical climates: Beijing, Berlin, Geneva, Incheon, London, Prague, Sapporo, Ulaanbaatar, Addis Ababa, Bangkok, Colombo, Harare, Havana, Lima, Phoenix, and Singapore. Weather data (EPW files) for each location are obtained from the EnergyPlus climate database [15, 16].

**Supplementary Figures**

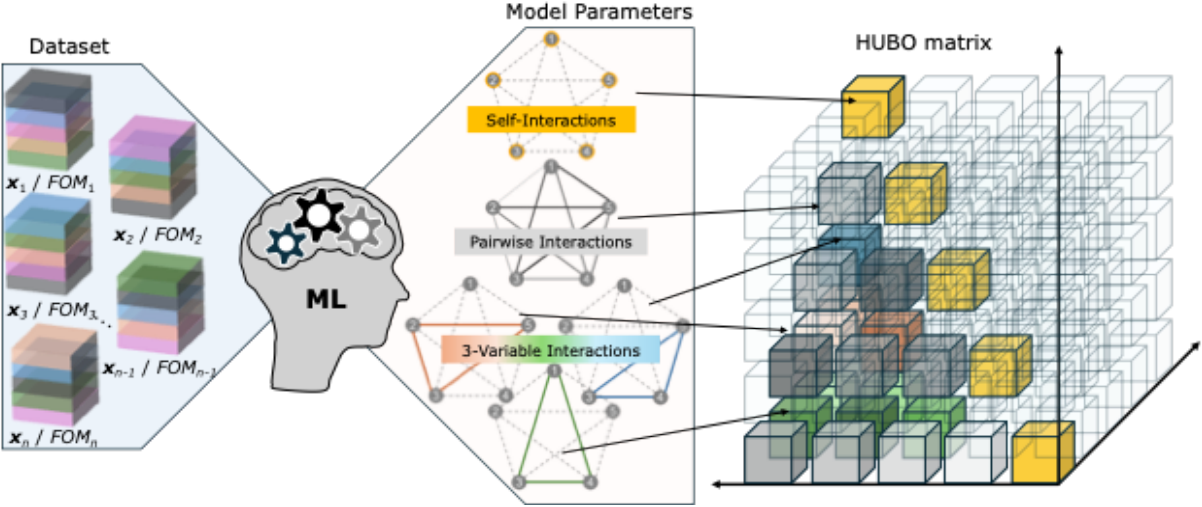

**Fig. S1: Schematic of higher-order FM surrogate modeling.** The trained FM generates linear, quadratic, and cubic coefficients, which are mapped to the HUBO representation used for optimization.

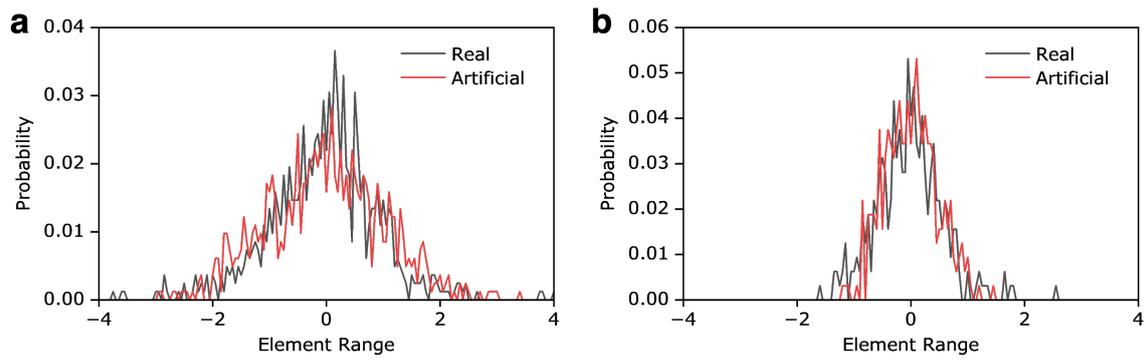

**Fig. S2: Distribution of HUBO Elements. a**, Distribution of linear, quadratic, and **b**, cubic coefficients obtained from the higher-order factorization machine (FM) surrogate model trained on representative real-world materials data. The empirical distributions are used to generate synthetic HUBO instances throughout this study.

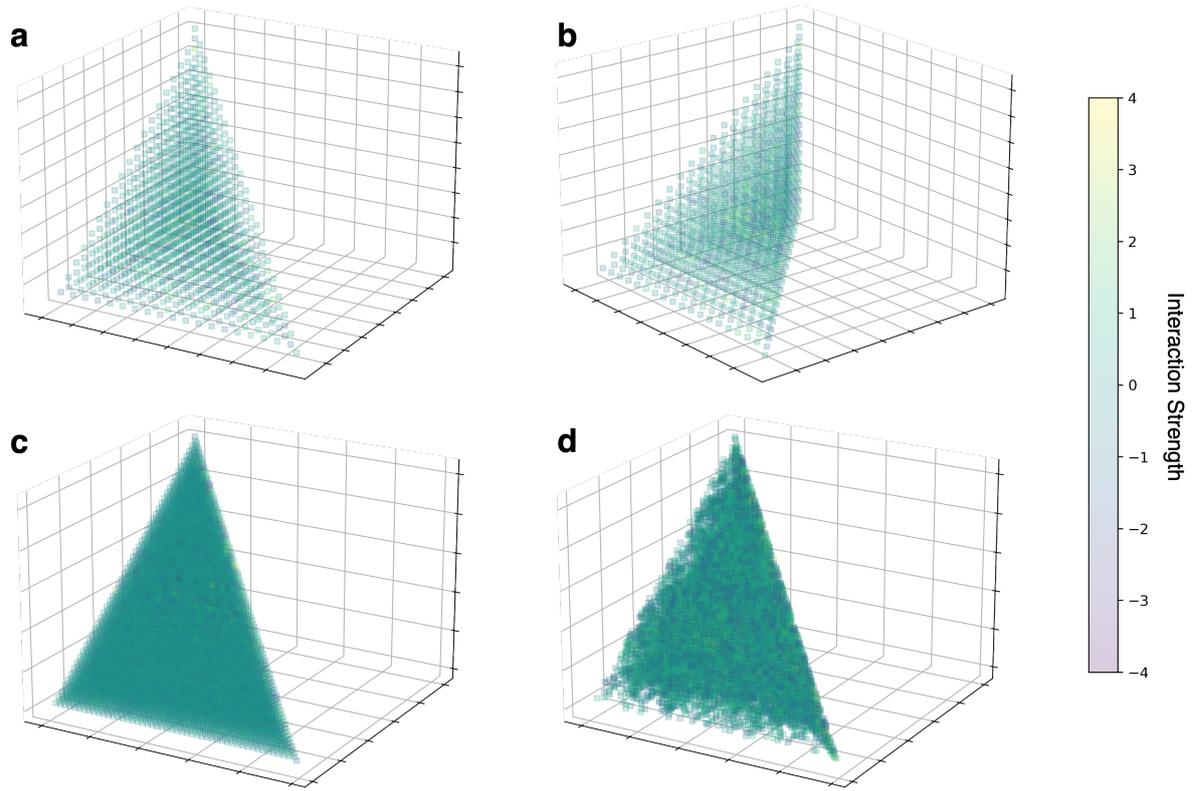

**Fig. S3: Structure of the HUBO matrices.** Three-dimensional visualization of HUBO matrices for problem sizes **a, b**, $N = 20$; **c**, $N = 50$; and **d**, $N = 100$. For clarity, elements with negligible magnitude (interaction strength $< 0.6$) are suppressed for $N = 100$. The HUBO matrices exhibits a triangular-pyramid structure, reflecting fully connected interactions across variables.

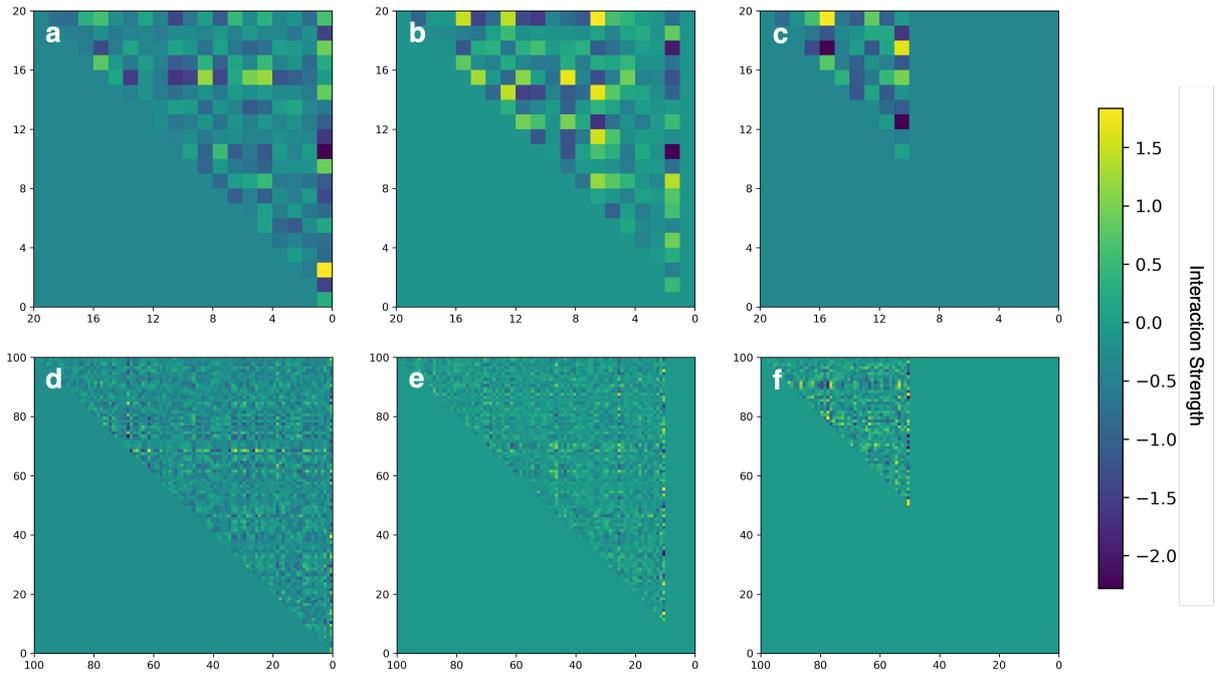

**Fig. S4: Selected planes of HUBO matrices.** Slices of the HUBO matrices at **a**, $0^{th}$; **b**, $1^{st}$; and **c**, $10^{th}$ index planes for $N = 20$, and at **d**, $0^{th}$; **e**, $10^{th}$; and **f**, $50^{th}$ index planes for $N = 100$. Each plane illustrates the dense (fully connected) interaction characteristic of higher-order FM–derived HUBO representations found in real-world (materials) optimization problems.

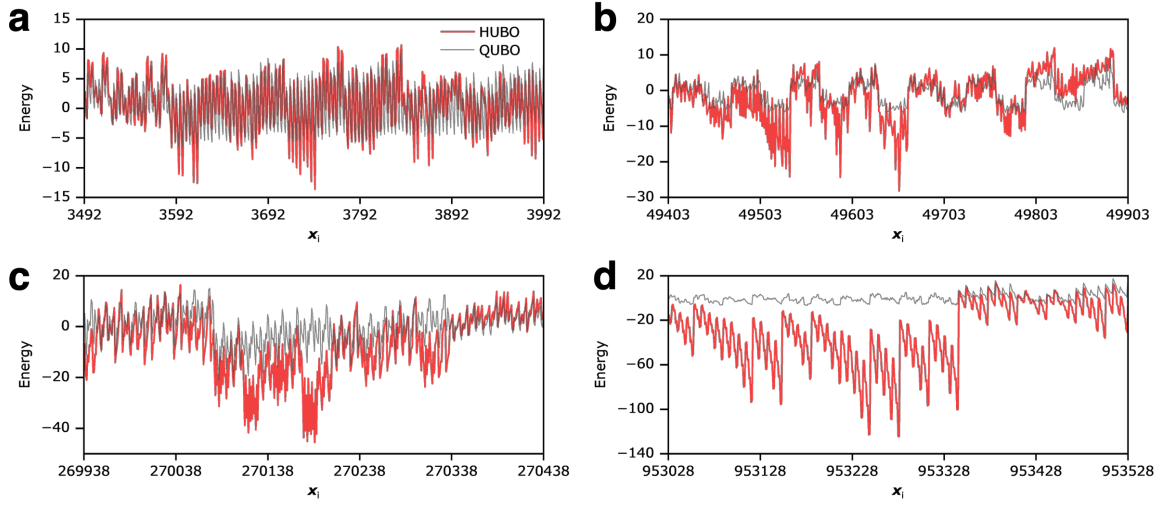

**Fig. S5: Rugged Energy Landscape of HUBO Problems.** Energy landscape of representative HUBO and QUBO instances for **a**, $N = 12$; **b**, $N = 16$; **c**, $N = 20$; and **d**, $N = 24$. The inclusion of three-variable interactions in HUBOs produces substantially more rugged landscapes compared to their quadratic (QUBO) counterparts, highlighting the increased difficulty of solving higher-order optimization problems.

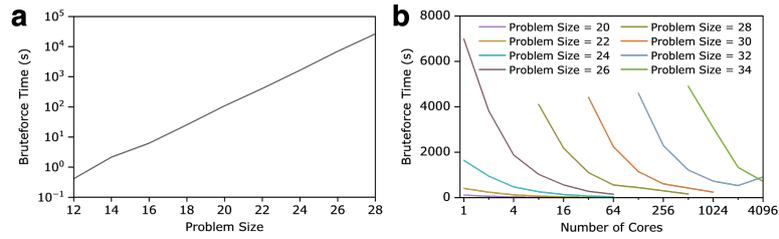

**Fig. S6: Brute-force scaling for HUBO problems. a**, Exponential growth of brute-force time with increasing problem size. **b**, Brute-force time under parallel execution on an high-performance computing system, illustrating that exact enumeration remains impractical even when employing massive parallelism.

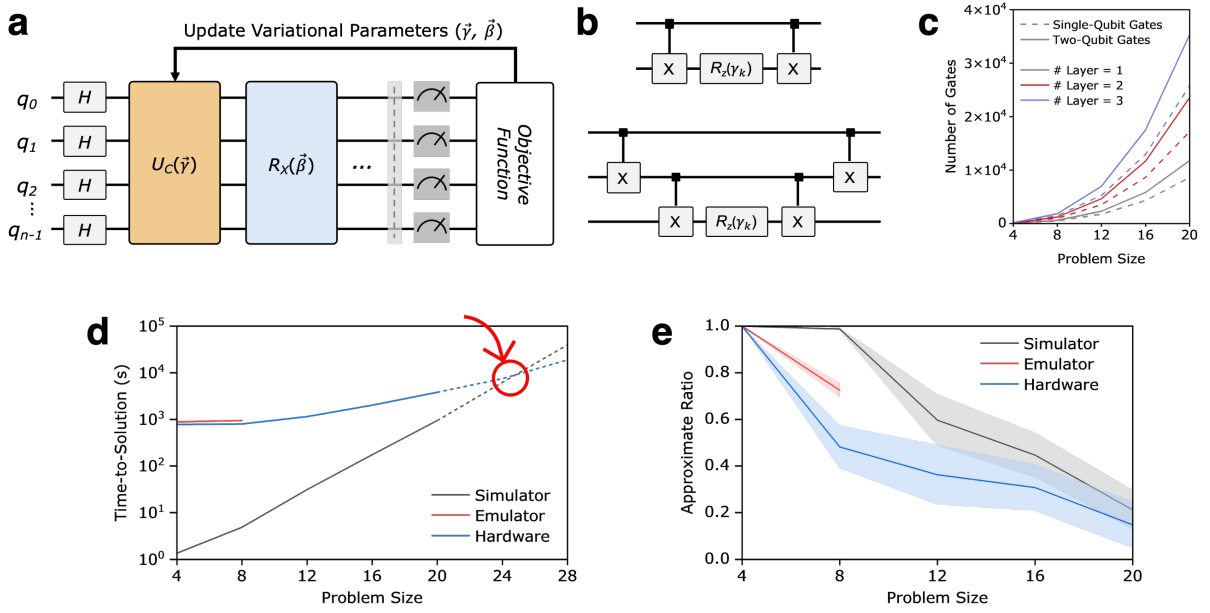

Fig. S7: **QAOA for HUBO problems. a**, Schematic of QAOA circuits by mapping pairwise and three-variable interactions via **b**, 2-local and 3-local terms. **c**, Scaling of QAOA circuit gate counts with problem size, showing the rapidly increasing circuit depth required to encode higher-order problems. **d**, Comparison of quantum simulator and hardware scaling behavior, highlighting the exponential cost of quantum simulator while the linear cost of hardware. The red arrow highlights the expected crossover point at which quantum hardware achieves faster execution than the simulation. **e**, Degradation of QAOA performance with increasing problem size, motivating the need for distributed approaches capable of addressing dense, large, complex HUBO problems. `Qiskit-aer`, IBM Heron r2 quantum processor (IBM-Kingston or IBM-Fez), and its corresponding emulator (fake backend) are used for simulator, hardware, and emulator, respectively.

11

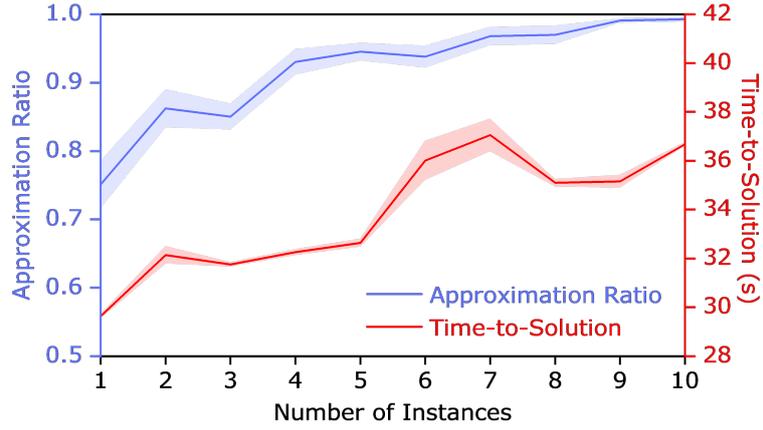

**Fig. S8: Scaling of DQOF with parallel instances.** Approximation ratio and time-to-solution of DQOF for a problem size $N$ of 40 as a function of the number of parallel instances $P$, illustrating improved solution quality with only marginal runtime increases as $P$ grows.

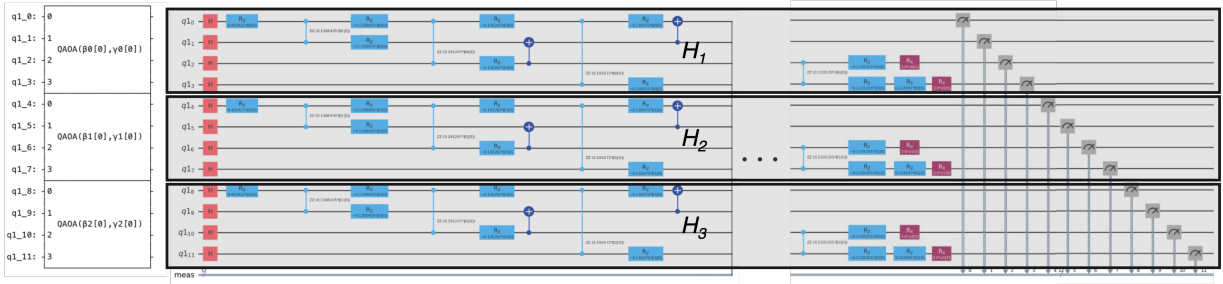

**Fig. S9: Construction of wide combined circuits via clustering.** Example quantum circuits generated by clustering multiple sub-HUBOs. As the number of clusters $m$ increases, circuit width expands while circuit depth remains constant, demonstrating strong compatibility with depth-limited near-term quantum hardware.

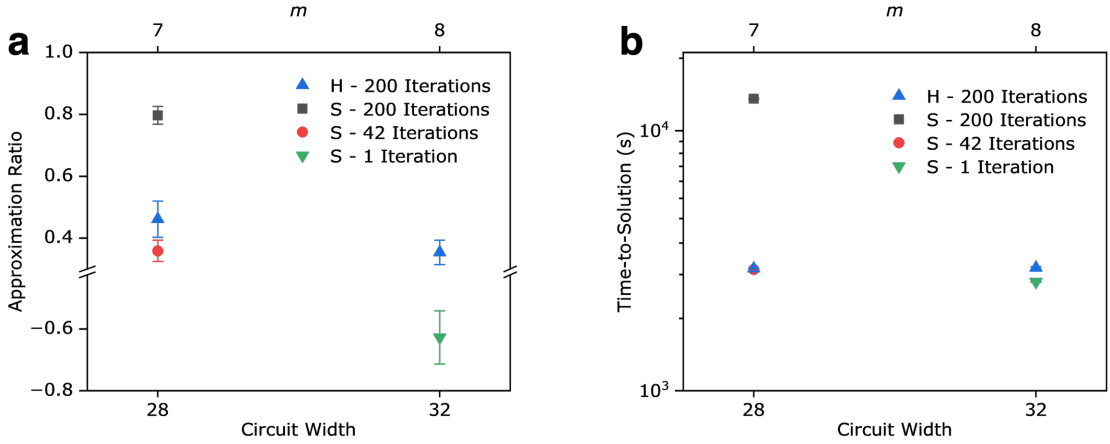

Fig. S10: **Performance of DQOF for combined Hamiltonians for the 'S' and 'H' configurations by clustering $m = 7$ and $m = 8$ sub-HUBOs. a**, Approximation ratio, and **b**, time-to-solution with different number of QAOA iterations for expectation value minimization. The number of iterations in the 'S' configuration is adjusted to yield a time-to-solution comparable to that of the 'H' configuration. Hardware execution maintains an approximately constant runtime even for wide circuits, whereas simulator-based execution achieves comparable time-to-solution only by substantially reducing the number of QAOA iterations. Such reductions, however, significantly degrade the approximation ratio, highlighting the advantage of hardware execution in achieving higher solution quality at similar runtimes.

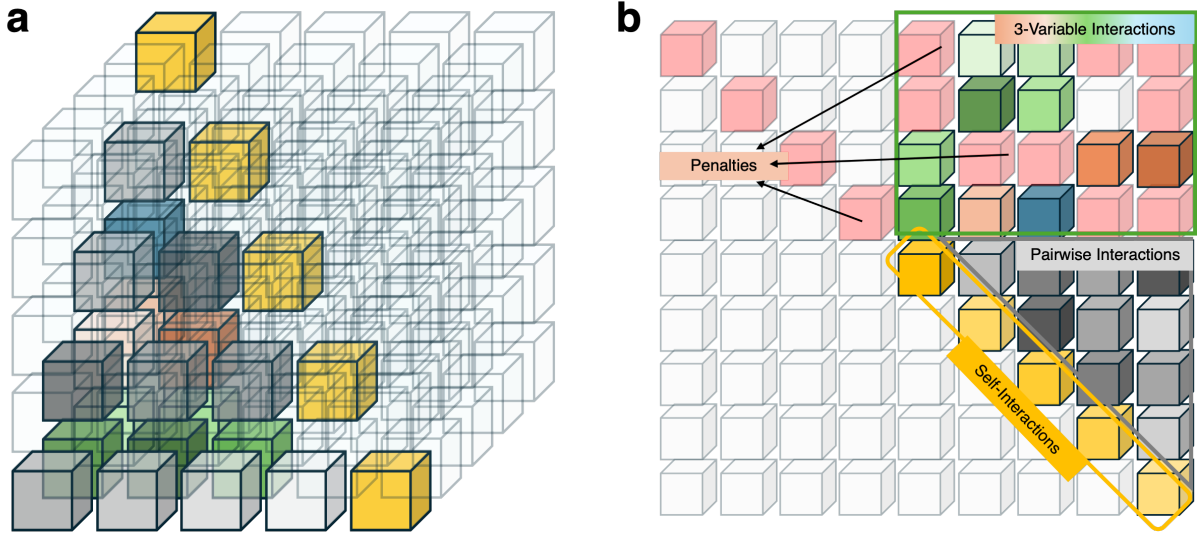

**Fig. S11: Quadratization of HUBOs into QUBOs. a**, Eaxmple HUBO matrix for $N = 5$. **b**, Resulting quadratized QUBO formulation, where auxiliary variables are introduced to encode three-variable interactions with penalty terms. Quadratization significantly increases problem dimensionality and alters the optimization landscape.

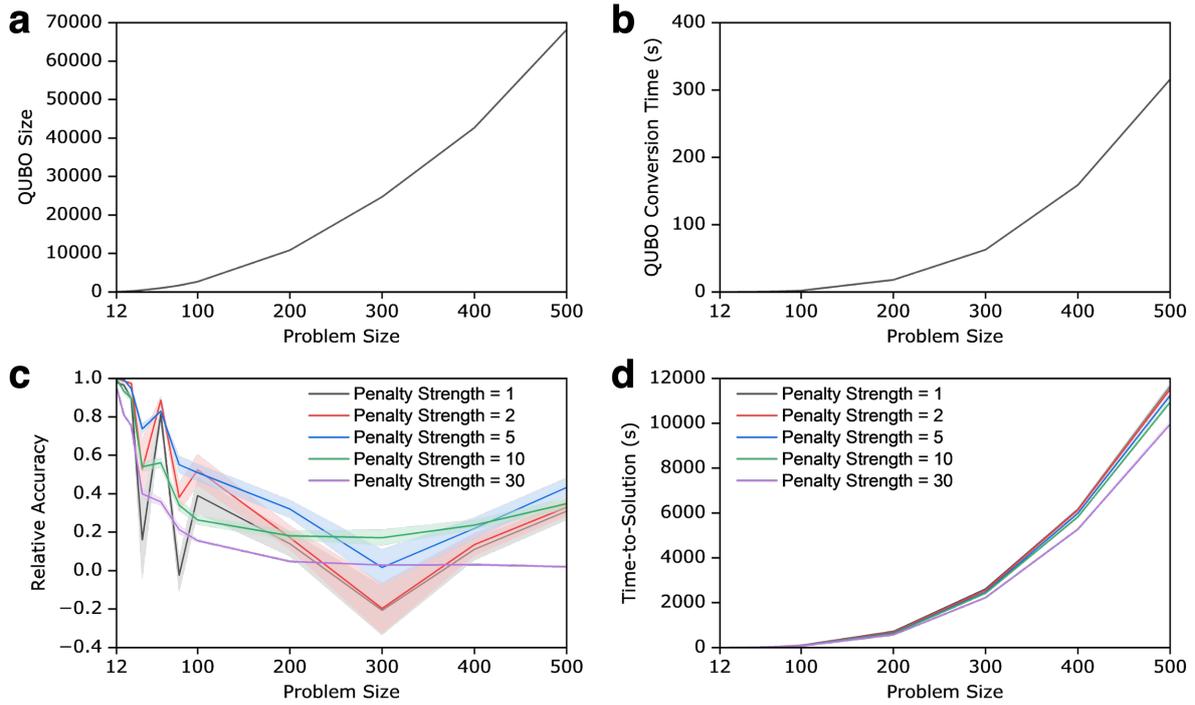

**Fig. S12: Growth of complexity of qadratized QUBOs. a**, Scaling of the resulting QUBO size as a function of the problem size. **b**, Runtime required to convert HUBOs into QUBOs. **c**, Relative accuracy and **d**, Time-to-solution of SA applied to quadratized QUBOs under different penalty strengths. These results show that quadratization performance is sensitive to the choice of penalty strength used to enforce auxiliary-variable constraints, and that no single penalty value is globally optimal across different problem sizes.

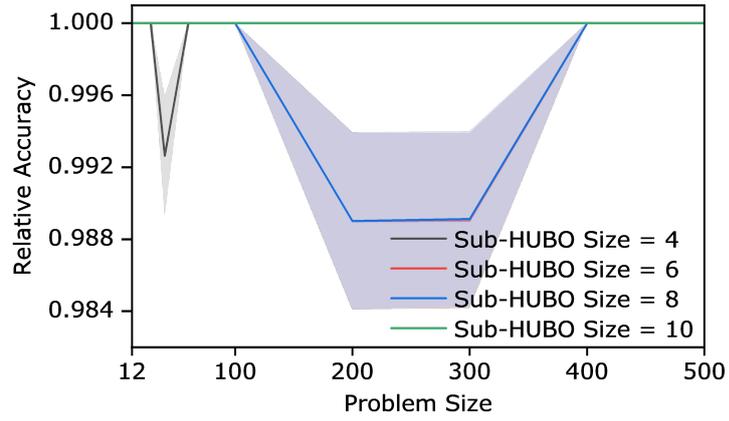

**Fig. S13: Effect of sub-HUBO size on DQOF performance.** Larger sub-HUBOs improve relative accuracy by capturing a greater portion of higher-order interactions within each DQOF iteration.

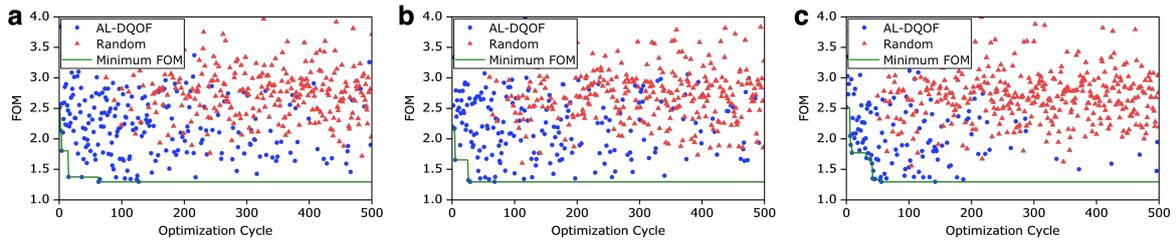

Fig. S14: **Figure of merit (FOM) distributions after TRC optimization using active learning with DQOF (AL-DQOF) for** $N = 12$. **a**, FOM distribution using the clustering strategy under the 'S+H' configuration. **b**, FOM distribution using the clustering strategy under the 'S' configuration. **c**, FOM distribution without the clustering strategy under the 'S' configuration. All configurations reliably converge toward the global optimal design, demonstrating the feasibility of AL-DQOF with the clustering strategy and hardware implementation for materials optimization.

| | S | S+H |
|---|---|---|
| Solar transmission efficiency | 0.7209 | 0.8224 |
| Solar reflection efficiency | 0.2791 | 0.1776 |
| Visible transmission efficiency | 0.8989 | 0.8378 |
| Visible reflection efficiency | 0.1011 | 0.1622 |
| IR transmission efficiency | 0.3906 | 0.3869 |
| Hemispherical emission efficiency | 0.5316 | 0.5256 |

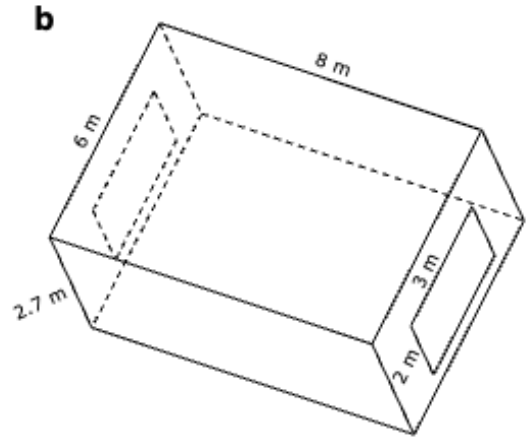

**Fig. S15: Energy-saving capability of optimized TRCs for window applications. a**, Optical properties of the TRCs optimized under the 'S' (40-bit system) and 'S+H' (12-bit system) configurations. **b**, Schematic of the small office model used in EnergyPlus simulations to evaluate cooling energy consumption. The optimized TRCs significantly reduce cooling load relative to standard glazing, demonstrating practical benefits of AL-DQOF–designed materials.



\end{document}